%% file: main.tex
\documentclass[11pt]{article}
\usepackage{geometry}                
\usepackage{graphicx}
\usepackage{amssymb}
\usepackage{epstopdf}
\usepackage{subfig}
\usepackage{amsmath}    
\usepackage{xcolor}
\usepackage[normalem]{ulem}

\usepackage{MnSymbol}	
\usepackage{natbib}
\usepackage{floatrow}
\usepackage{authblk}

\newcommand{\metrict}{\boldsymbol{\mathrm{g}}}
\newcommand{\vect}[1]{\boldsymbol{#1}}
\newcommand{\vectn}[1]{\boldsymbol{\hat{#1}}}
\newcommand{\grad}[1]{\boldsymbol{\nabla}#1}
\newcommand{\diverg}[1]{\boldsymbol{\nabla}\cdot #1}

\title{Internal energy dissipation in Enceladus's ocean from tides and libration \& the role of inertial waves}
\author[1]{J. Rekier, A. Trinh, S. A. Triana, V. Dehant}
\affil[1]{Royal Observatory of Belgium}
\date{\today}

\begin{document}
\maketitle

\begin{abstract}
Enceladus is characterised by a south polar hot spot associated with a large outflow of heat, the source of which remains unclear. We compute  the viscous dissipation resulting from tidal and libration forcing in the moon's subsurface ocean using the linearised Navier-Stokes equation in a 3-dimensional spherical model. We conclude that libration is the dominant cause of dissipation at the linear order, providing up to $\sim0.001$~GW of heat to the ocean, which remains insufficient to explain the $\sim10$~GW observed by Cassini. We also illustrate how resonances with inertial modes can significantly augment the dissipation. Our work is an extension to \citet{Rovira-Navarro2019} to include the effects of libration. The model developed here is readily applicable to the study of other moons and planets.
\end{abstract}

\input{introduction.tex}

\input{method.tex}

\input{results.tex}

\input{discussion.tex}

\input{acknowledgement.tex}

\bibliographystyle{abbrvnat}
\bibliography{bibliography}

\input{appendix.tex}

\end{document}

%% file: introduction.tex
\section{Introduction}

From the massive amount of information collected by Cassini, Saturn's moon Enceladus currently appears as one of the most habitable moons in the Solar System. The presence of a subsurface reservoir of liquid water was inferred soon after the first few flybys \citep{Porco2006}. Measurements of the gravity field deduced from subsequent flybys helped establish the large spatial extent of this reservoir \citep{Iess2014}, but only libration observations could provide definitive evidence for a global-scale ocean rather than a regional sea \citep{Thomas2016}. 

Tidal dissipation is the most likely power source for the observed geological activity \citep{Nimmo2018}. Current models, however, have so far been unable to account for the observed $\sim10$~GW of endogenic heat flow emanating from the south polar region \citep{Spencer2006,Howett2011} without invoking the presence of a highly porous (`fluffy') solid inner core \citep{Roberts2015,Choblet2017} or a convecting icy crust \citep{Behounkova2010,Behounkova2017}. As suggested by \citet{Barr2007}, this latter hypothesis appears unlikely in view of the relatively small thickness of the crust ($\sim 20$~km) deduced from geodetic observations \citep{Beuthe2016}.

If dissipation is not concentrated in the core or in the shell, one possibility is that it takes place predominantly in the ocean. Most of the current models focusing specifically on the ocean layer rely on the solution of the Laplace Tidal Equations (LTE) whereby the fluid is modelled as a 2-dimensional thin layer \citep{Tyler2014,Hay2017,Beuthe2016,Matsuyama2018}. However, gravity and topography data suggest that the average thickness of the ocean is not negligible, $\sim38$ km if the crustal topography is isostatically supported, which is also consistent with the thin crust inferred from libration \citep{Beuthe2016,Hemingway2019}.
Such a thick ocean challenges the validity of the conclusions drawn from the 2-d models. 

Additionally, the action of the Coriolis force on planetary oceans is known to support the existence of inertial waves within their volume \citep{Poincare1885}. \citet{Morize2010} have proven that inertial waves can be excited through tidal deformation in their laboratory experiment. The role they play in planetary and astrophysical settings is still, however, far from clear. When viscosity is taken into account, the flow associated with these waves develops regions of intense shear within the fluid volume \citep{greenspan1968}. These \emph{internal shear layers} can significantly increase the total amount of dissipation in the ocean. The formula giving the total viscous dissipation for a flow of velocity field $\vect{v}$, is 
\begin{equation}
\mathcal{D}_\text{visc}=2~\text{Ek}\int_\mathcal{V}\mathord{\stackrel{\smallfrown}{\grad{\vect{v}}}}:\mathord{\stackrel{\smallfrown}{\grad{\vect{v}}}}~,
\label{eq:Dvisc}
\end{equation}
with $\mathord{\stackrel{\smallfrown}{\grad{\vect{v}}}}\equiv\frac{1}{2}(\grad{\vect{v}}+\grad{\vect{v}}^\text{T})$ and where $\text{Ek}$ denotes the Ekman number, which is typically very small in planetary applications ($\text{Ek}\sim10^{-13}$ for Enceladus). In order to significantly contribute to the total dissipation, the gradient of velocity within the ocean volume in general and in the internal shear layers in particular, must be sufficiently strong to compensate the smallness of $\text{Ek}$. A study of the problem has recently been conducted by \citet{Rovira-Navarro2019} who concluded to the irrelevance of inertial waves in the energy budget. We reproduce their results in the present paper and extend them to include the contribution from libration which corresponds to a forcing with a much larger amplitude. We also take into account the presence of the icy crust. As the obliquity of Enceladus is very small \citep{Baland2016}, we focus our attention on the effects of eccentricity tides, which are likely to dominate over obliquity tides in the ocean layer \citep{Chen2011,Tyler2014}, even though, recent work by \citet{Hay2019} has shown that non-linear effects can alter this picture at least within the simplified model based on the LTE. 

The present paper is structured as follows. In Sec.~\ref{sec:method}, we present our model including the details on the equilibrium state and the equations of motion and how these are used to compute the response to tidal and libration forcing. The results are presented in Sec.~\ref{sec:results} and discussed in Sec.~\ref{sec:discussion}. We provide details about the interpretation of the forced response in terms of resonance with inertial waves in this last section and conclude with perspectives about the possible continuations of the present work.

The generality of the method described below makes it readily applicable to the study of other moons and planets.

%% file: method.tex
\section{Method}
\label{sec:method}

We can only reach the very low values of $\text{Ek}$ in a sphere as it is not otherwise possible to model the very thin viscous layer at the ocean's boundary.
For this reason, we proceed in two steps. First we compute the tidal deformation at the ocean's boundary of a non-rotating spherical model (Sec. \ref{sec:deformation}). Second, we compute the resulting flow inside the ocean and with rotation taken into account (Sec.~\ref{sec:NS}).

\subsection{Deformation}
\label{sec:deformation}
\subsubsection{Equilibrum state}
We model the moon as a set of 3 homogeneous concentric spherical shells and we assume that each layer is incompressible. Table~\ref{tab:parameters} gives the set of parameters used in the present paper. We set the size of the core to the central value 192 km obtained by \citet{Beuthe2016} with the reference datasets, and consider a range of interior models with variable ocean thickness.
\begin{table}[ht]
\caption{List of physical parameters} 
\centering 
\begin{tabular}{| l | c | c c c |} 
\hline 
 Name & Symbol & Core & Ocean & Crust \\
\hline\hline 
 Outer radius (km) &R & 192 & [192 - 252] & 252 \\
 Mass Density (kg/$\text{m}^3$) & $\rho$ & [2483 - 2357]* & 1020 & 920 \\
 Shear modulus (GPa) & $\mu$ & 40 & 0 & 3.5 \\
 \hline
\end{tabular}\\
\label{tab:parameters} 
\small{* the density of the core is adjusted to conserve the total mass ($M=1.080\times10^{20}$kg)}
\end{table}

In what follows, we use $\eta=\frac{R_\text{core}}{R_\text{ocean}}$ to represent the aspect ratio of the spherical shell.

\subsubsection{Equations of deformation}
The equation governing the oscillation around the equilibrium state are the Poisson equation and the conservation of momentum. Assuming incompressibility, these read :
\begin{align}
\nabla^2\delta\phi&=0~,\label{eq:poisson}\\
-\omega^2\rho\vect{s}&=\diverg{\vect{\Delta\sigma}}-\grad{(\rho\vect{s}\cdot\grad{\phi}_0)}-\rho\grad{\delta\phi}\label{eq:momentumeq}~,
\end{align}
where $\phi_0$ is the equilibrium gravity potential, $\delta\phi$ denotes the (Eulerian) increments of gravity (\emph{i.e.} self-gravitational + tidal) potential, $\vect{s}$ is the displacement vector field and $\vect{\Delta\sigma}$ denotes the (Lagrangian) increment of Cauchy stress~\citep{dahlen1998}. The latter can be expressed in terms of the displacement and (Lagrangian) increment of pressure~:
\begin{equation}
\vect{\Delta\sigma}\equiv-p\metrict+\mu\left(\grad{\vect{s}}+\grad{\vect{s}}^\text{T}\right)~,
\end{equation}
where $\metrict$ is the metric tensor. $\mu$ denotes the shear modulus, which is assumed to be constant in each layer, its values are listed in Table~\ref{tab:parameters}, as well as those for the density, $\rho$.

The gravity potential, displacement vector and stress tensor must satisfy the following conditions at each undeformed spherical boundary~\citep{dahlen1998}~:
\begin{align}
\left[\delta\phi\right]^+_-&=0~,\\
\left[\vectn{n}\cdot\left(\grad{\delta\phi}+4\pi G\rho\vect{s}\right)\right]^+_-&=0~,\\
[\vectn{n}\cdot\vect{\Delta\sigma}]^+_-&=0~,
\end{align}
where $\vect{\hat{n}}$ is the normal vector and the notation $\left[\cdot\right]^+_-$ denotes the difference between the values of the enclosed quantity on both sides of the boundary. The last constraint is that the normal component of the displacement be continuous across each internal boundary :
\begin{equation}
[\vectn{n}\cdot\vect{s}]^+_-=0~.
\end{equation}
The above does not hold at the free outermost boundary.

\subsubsection{Tidal deformation}
Enceladus gets deformed under the effect of tides caused by its slightly eccentric orbit around Saturn (eccentricity tides). The corresponding (Eulerian) increment of gravity potential can be decomposed in terms of spherical harmonics :
\begin{equation}
\delta\phi=\sum_{\ell=0}^\infty\sum_{m=-\ell}^{\ell}\delta\phi_{\ell,m}\text{Y}_\ell^m(\theta,\lambda)~,
\end{equation}
where $\theta$ and $\lambda$ denote the colatitude and (east) longitude respectively.\footnote{We use the convention that $\text{Y}_\ell^m(\theta,\lambda)\equiv\sqrt{\frac{(\ell-m)!}{(\ell+m)!}}\text{P}_\ell^m(\cos\theta)e^{im\lambda}$ \\ and $\text{P}_\ell^m(x)=\frac{(-1)^m}{2^\ell\ell!}(1-x^2)^\frac{m}{2}\frac{d^{\ell+m}}{dx^{\ell+m}}(x^2-1)^\ell$, where $\text{P}_\ell$ are the Legendre polynomials.}
The only nonzero components for the contribution due to tides caused by the eccentricity of Enceladus' orbit are
\begin{align}
\delta\phi^\text{tide}_{2,-2}&=-\frac{7}{4} \sqrt{\frac{3}{2}} e \omega ^2 r^2\\
\delta\phi^\text{tide}_{2,0}&=\frac{3}{4} e \omega ^2 r^2\\
\delta\phi^\text{tide}_{2,2}&=\frac{1}{4} \sqrt{\frac{3}{2}} e\omega ^2 r^2 ~,
\end{align}
where $\omega=\frac{2\pi}{1.37}~\text{days}^{-1}$ is the (diurnal) frequency of the forcing and $e=4.7\times10^{-3}$ denotes the orbital eccentricty.

We compute the resulting displacement at the top and bottom of the ocean using Eq.~(\ref{eq:poisson}) \& (\ref{eq:momentumeq}) for different values of $\eta$. More details are given in Appendix~\ref{sec:tidaldisplacements}. Fig.~\ref{fig:tidaldeformation} shows the spatial pattern for each tidal component for $\eta=0.838$ corresponding to a shell thickness of 23 km, as suggested by models of isostasy \citep{Beuthe2016}. 
\begin{figure}
\captionsetup[subfloat]{labelformat=empty}
\centering
\begin{tabular}{ccc}
\subfloat[$m=-2$]{\includegraphics[width=0.31\textwidth]{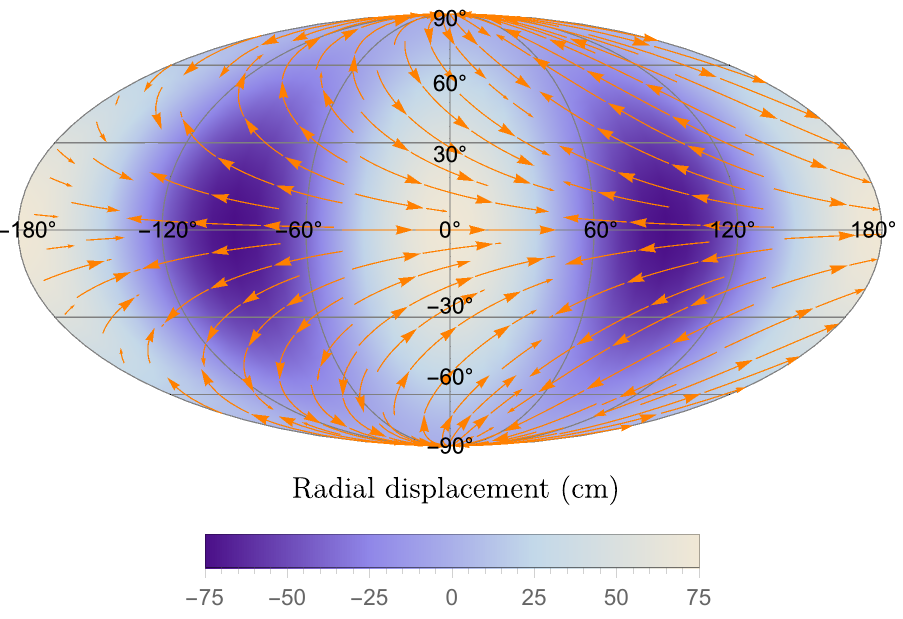}} &
\subfloat[$m=0$]{\includegraphics[width=0.31\textwidth]{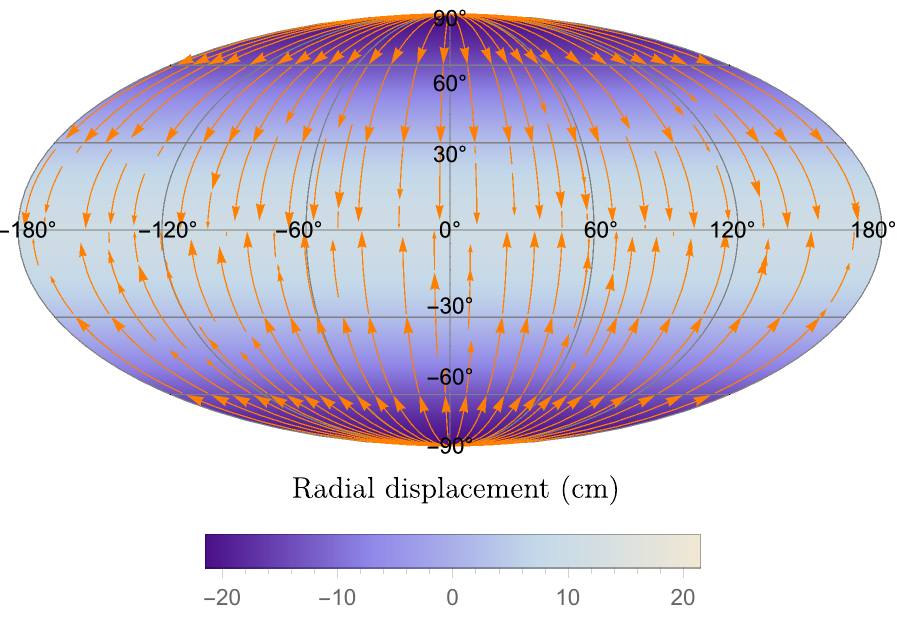}}&
\subfloat[$m=2$]{\includegraphics[width=0.31\textwidth]{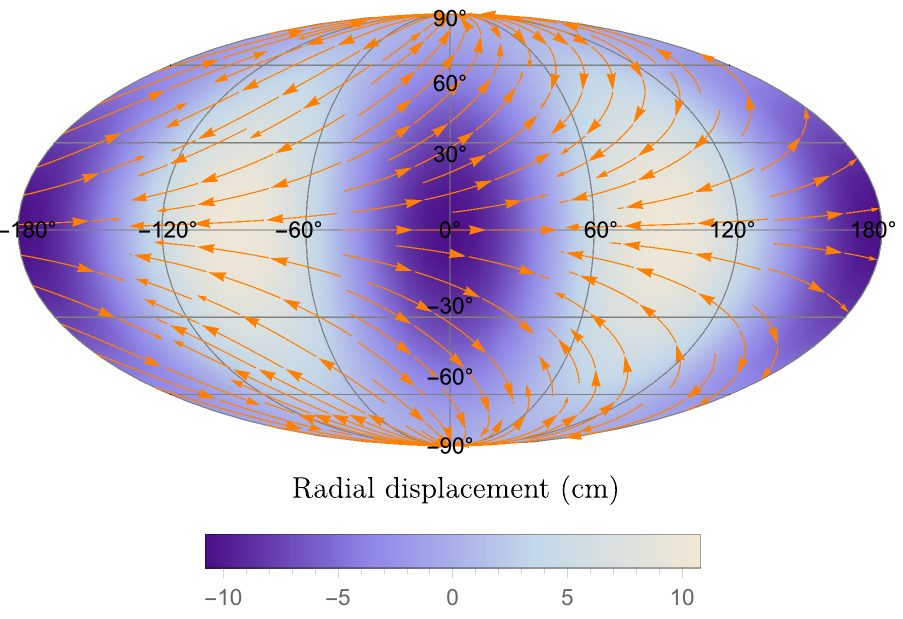}}
\end{tabular}
\caption{Deformation pattern at the top of the ocean caused by the eccentricity tides on Enceladus (ocean aspect ratio $\eta=0.838$). The arrows indicate the tangential displacement and are not to scale.}
\label{fig:tidaldeformation}
\end{figure}

\subsubsection{Libration forcing}
Enceladus is not, in reality, spherical. And this fact influences its rotation. In particular, the equilibrium deformation caused by permanent tides induces a periodic motion of longitudinal libration under the perturbing effect of gravitational torques from external sources and the restoring effect of internal pressure and gravitational torques (inertial torques). 

In the present paper, we treat the libration motion as a given and compute the resulting dissipation in the fluid layer of our simplified spherical setting. Our model of libration is based on the rigid-shell model of \citet{Baland2010}; we therefore neglect the limited decrease of libration amplitude introduced by elastic deformation \citep{VanHoolst2016}.
For the present, it is sufficient to observe that the libration of the ocean's flattened boundaries can be represented as a superposition of several oscillations of fictitious spherical boundaries: a toroidal degree-1 oscillation where the librating spherical boundaries tend to viscously drag the ocean, and a spheroidal degree-2 oscillation where the librating flattened boundaries tend to push the ocean fluid, as if spherical boundaries underwent radial deformation. 

The tangential displacement can be obtained directly from the libration amplitude as a function of $\eta$, the aspect ratio of the ocean (see Fig.~\ref{fig:libration_amplitude} of Appendix~\ref{sec:librationforcing}). It results in what is later referred to as the $m=0$ component. 

The non-axisymmetric shape of Enceladus also causes a radial displacement at the top and bottom of the ocean, which can be estimated in the following manner. To first order, the radial coordinate of the triaxial boundary can be approximated as 
\begin{equation}
r(\theta,\lambda)=R\{1+\alpha_0 \text{Y}_2^0(\theta,\lambda)+\alpha_2 [\text{Y}_2^2(\theta,\lambda)+\text{Y}_2^{-2}(\theta,\lambda)]\}~,
\label{eq:triaxialshape}
\end{equation}
where $R$ is the mean radius and $\alpha_0$ and $\alpha_2$ are real numbers describing the polar and equatorial flattening, which we assume to be hydrostatic and compute using the first-order theory of figures. 
In the frame rotating at constant angular velocity $\vect{\Omega}$, libration can be accounted for by replacing : $\lambda\rightarrow\lambda+\delta(t)$ with $\delta\ll1$. The first two terms in Eq.~(\ref{eq:triaxialshape}) do not depend on the azimuthal angle, $\lambda$, and so the velocity of the moving boundary comes out to be 
\begin{equation}
\frac{\dot{r}}{R}=2 i \alpha_2\dot{\delta}[\text{Y}_2^2(\theta,\lambda)-\text{Y}_2^{-2}(\theta,\lambda)]~.
\end{equation}
Now, if we set $\delta(t)=\epsilon\sin(\omega t)$, where $\epsilon$ and $\omega$ denote the amplitude and the (diurnal) frequency of libration respectively, we obtain
\begin{equation}
\frac{\dot{r}}{R}=2\alpha_2 \epsilon (i\omega)[\text{Y}_2^2(\theta,\lambda)-\text{Y}_2^{-2}(\theta,\lambda)]\cos(\omega t)~.
\label{eq:libforcingrad}
\end{equation}
The values of the (hydrostatic) flattening parameter $\alpha_2$ and the amplitude of libration $\epsilon$ both depend on the aspect ratio $\eta$. We provide simplified expressions of these in Appendix~\ref{sec:librationforcing}. Hereafter, the radial displacement is referred to as the $m=\pm2$ components.

\subsection{Viscous dissipation}
\label{sec:NS}

We use the (linearised) Navier-Stokes equation to model the motion of the ocean. In its dimensionless form and in the frame rotating with angular velocity $\vect{\Omega}~\equiv~\Omega~\vect{\hat{z}}$, this reads~:
\begin{equation}
i\omega\vect{v}+2\vect{\hat{z}}\times\vect{v}+\grad{p}-\text{Ek}\nabla^2\vect{v}=0~,
\label{eq:NS}
\end{equation}
where $\vect{v}$ denotes the velocity and $p$ denotes the reduced pressure. The Ekman number parametrises the balance between the viscous force and the Coriolis force. We define it as such
\begin{equation}
\text{Ek}\equiv\frac{\nu}{\Omega R_o^2}~,
\end{equation}
where $\nu$ denotes the kinematic viscosity (here taken to be that of water: $\nu=10^{-6}~\text{m}^2/\text{s}$) and $R_o$ is the (mean) radius at the top of the ocean. In planetary settings, Ek is typically very small. For Enceladus, its value is of the order\footnote{There is actually a lot of uncertainty on the precise value of this parameter. Here we are only interested in its order of magnitude.} $\text{Ek}\sim3.5\times10^{-13}$. 

The motion inside the ocean is forced by the deformation of the boundary caused by tides and libration. We recover the velocity of the moving boundary from the displacement vector $\vect{s}$ via~
\begin{equation}
\vect{v}_\text{tide/libration}=i\omega\vect{s}_\text{tide/libration}~.
\end{equation}
We use the \emph{no-slip} boundary condition, which, in this case, amounts to impose that the velocity is continuous across the boundary (at the top and bottom of the ocean)~:
\begin{equation}
\vect{v}=\vect{v}_\text{tide/libration}~.
\label{eq:bcv}
\end{equation}
To first order, it is sufficient to enforce this condition at the boundary of the (spherical) equilibrium figure. 
Owing to the condition of incompressibility ($\vect{\nabla}\cdot\vect{v}=0$) we can write the velocity field as
\begin{equation}
\vect{v}=\vect{\nabla}\times\vect{\nabla}\times(P\vect{r})+\vect{\nabla}\times(T\vect{r})~,
\label{eq:poltor}
\end{equation}
where $P$ and $T$ denote the poloidal and toroidal potentials respectively. These are then decomposed in terms of their spherical harmonics coefficients, $P_{\ell,m}$ and $T_{\ell,m}$. The two orthogonal projections of Eq.~(\ref{eq:NS}) that we use are given in Appendix~\ref{sec:NSYlm} in terms of their spherical harmonics projections as Eqs.~(\ref{eq:curlcurlNS}) and (\ref{eq:curlNS}). 

We solve these equations numerically using the techniques presented in Sec.~2 of \citet{Rekier2018}
and we compute the dissipation using Eq.~(\ref{eq:Dvisc}).

%% file: results.tex
\section{Results}
\label{sec:results}
\subsection{Tides}

\begin{figure}
\centering
\makebox[\textwidth][c]{\includegraphics[width=1.2\textwidth]{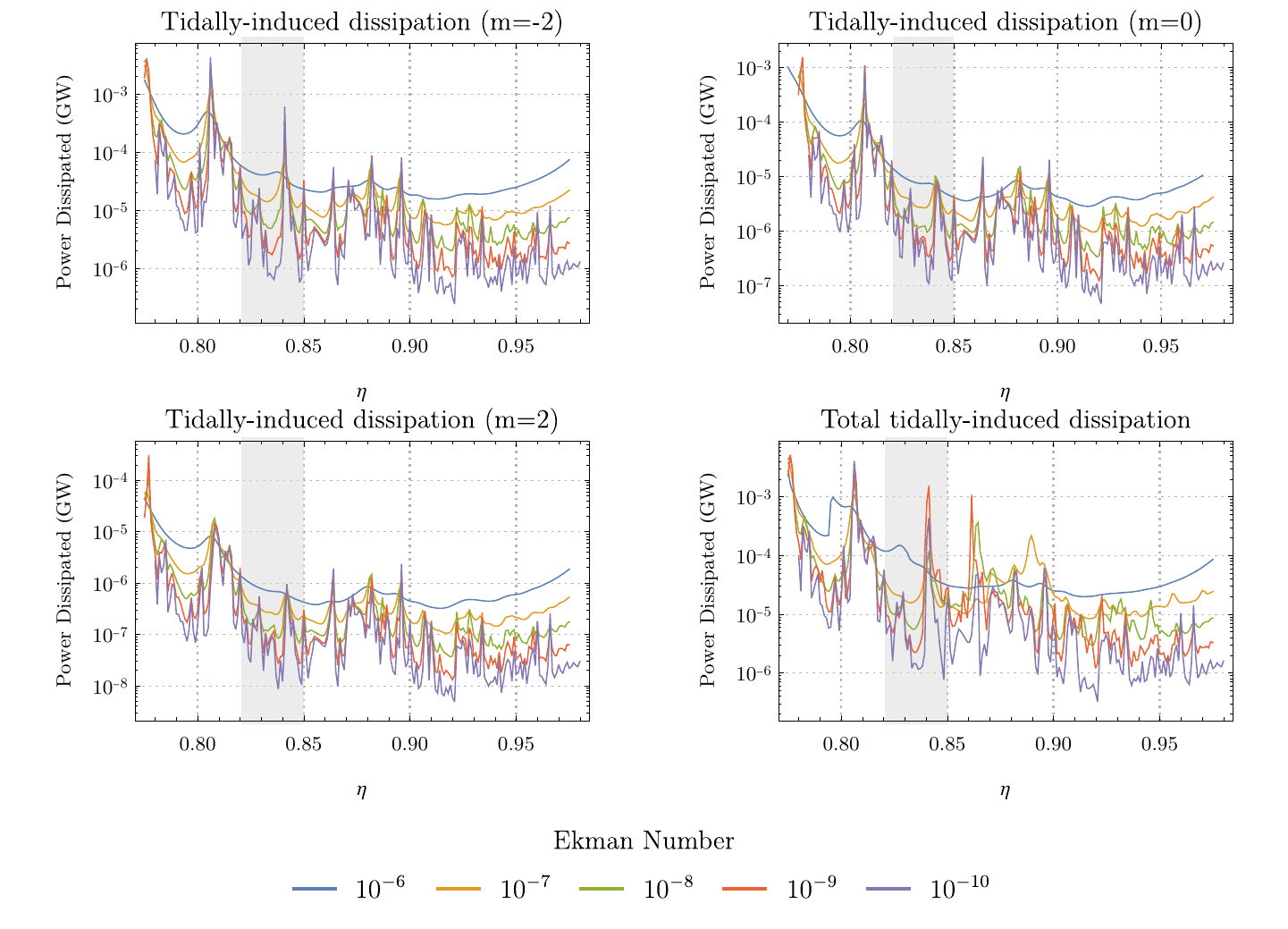}}%
\caption{{\bf Eccentricity tides} -- Viscous dissipation in Enceladus's ocean as a function of $\eta$, the aspect ratio of the spherical shell, for different values of the Ekman number. The bottom-right plot is the sum of all the others. The shaded area represents the range $0.82\leq\eta\leq0.85$ predicted by models of isostasy.}
\label{fig:tidaldissipation}
\end{figure}

Fig.~\ref{fig:tidaldissipation} shows the dissipation for $m\in\{-2,0,2\}$ and the total dissipation for all these contributions (lower-right panel). There is a general trend towards larger dissipation as $\eta$ decreases. This is consistent with the results of \citet{Matsuyama2018} as this regime corresponds to a thicker ocean and a thinner icy crust in our model. The principal contribution to the total dissipation comes from the prograde tide ($m=-2$) as expected considering it is the dominant contribution to the tidal potential. The total contribution from the eccentricity tides lies somewhere in the range $10^{-6} <\mathcal{D}_\text{visc}<10^{-3}$ GW depending on the presence or absence of a peak.

\begin{figure}
\centering
\begin{tabular}{c c}
\subfloat[$\eta=0.838$]{\includegraphics[width=0.50\textwidth]{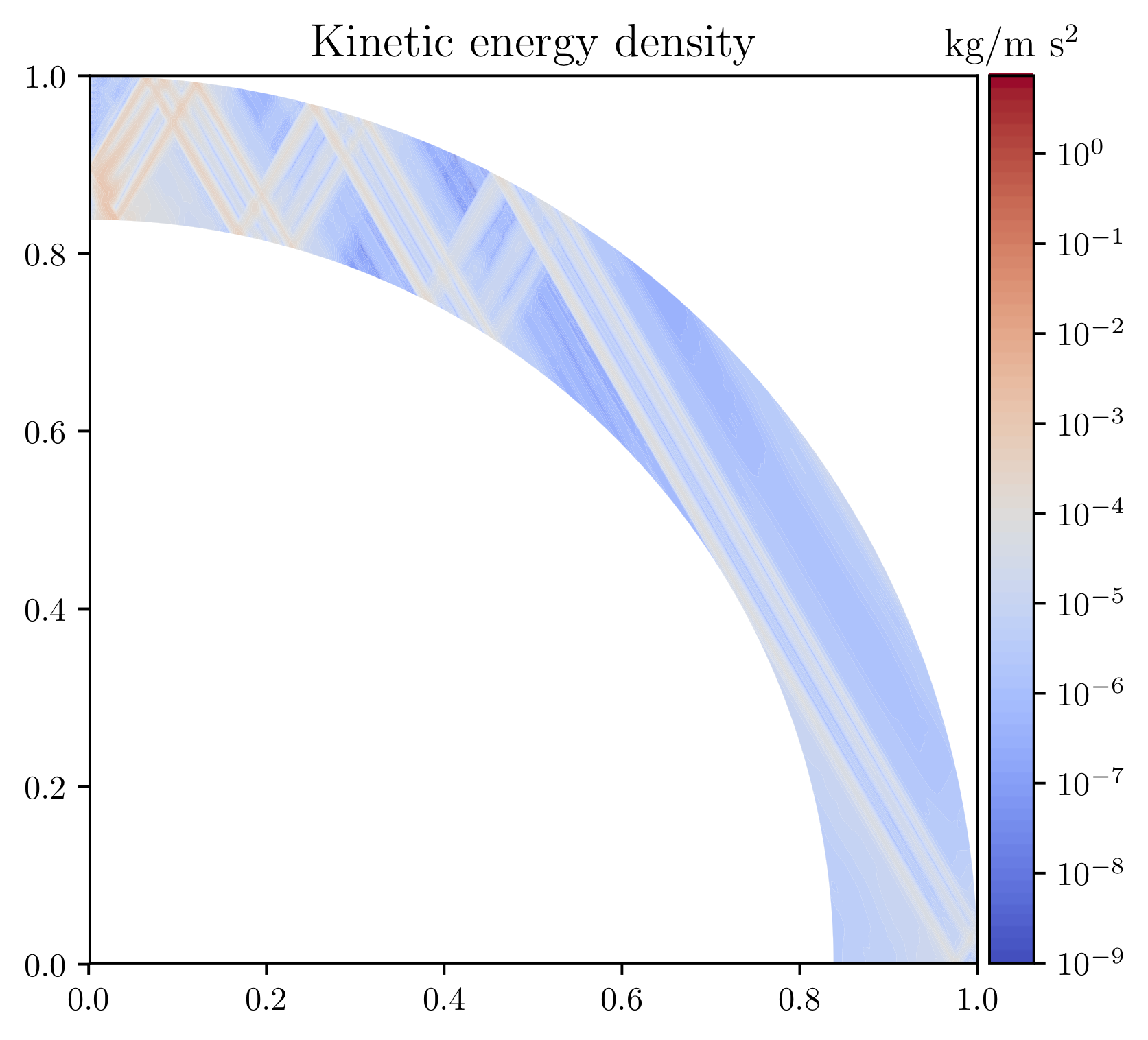}\label{fig:sla}}&
\subfloat[$\eta=0.841$]{\includegraphics[width=0.50\textwidth]{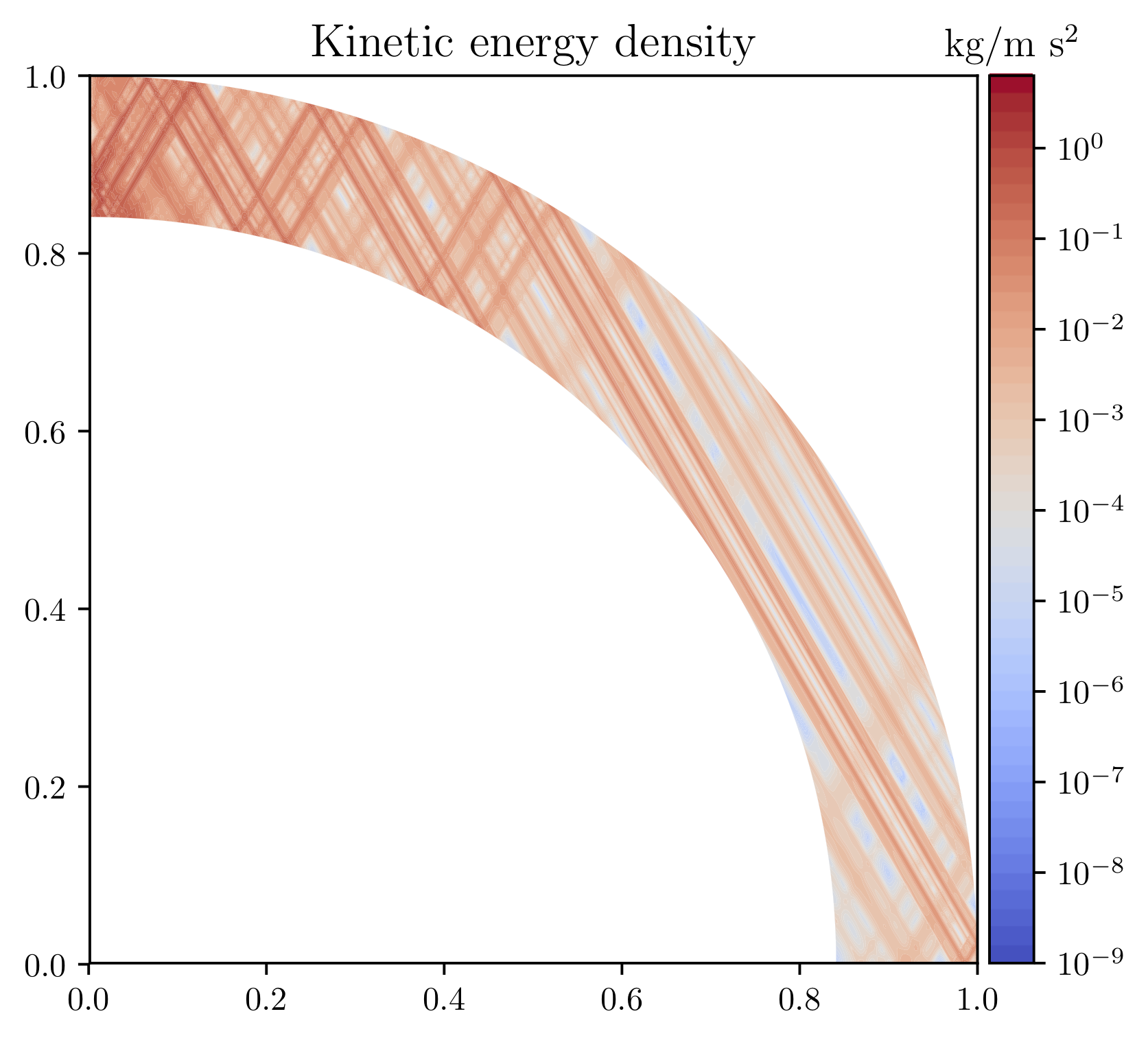}\label{fig:slb}}
\end{tabular}
\subfloat[]{\includegraphics[width=0.6\textwidth]{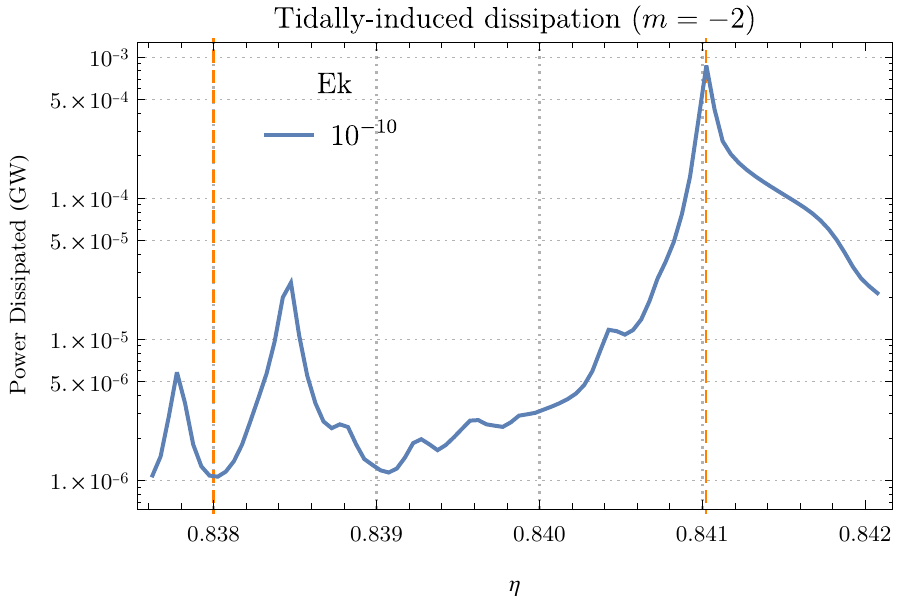}\label{fig:tide_D_zoom}}
\caption{{\bf Eccentricity tides} -- (a) and (b) : Density of kinetic energy inside the ocean for two values of the aspect ratio, $\eta$, close to its actual value for Enceladus ($m=-2$, $\text{Ek}=10^{-10}$). The associated amounts of dissipation for each case is shown on (c) and differ by three orders of magnitude.}
\label{fig:shearlayers}
\end{figure}

The peaks in the dissipation profile are due to the augmented contribution caused by the internal shear layers inside the ocean. This can be observed from the plots on Fig.~\ref{fig:sla}~\&~\ref{fig:slb}, which show the density of kinetic energy for two different values of $\eta$ (both within the the range $0.82<\eta<0.85$ obtained by \citet{Beuthe2016}) one associated to larger dissipation and one to smaller dissipation. The corresponding values of $\mathcal{D}_\text{visc}$ can be read from Fig.~\ref{fig:tide_D_zoom} and differ by three orders of magnitude. The pattern of internal shear layers appears much sharper on Fig.~\ref{fig:slb}, corresponding to the higher dissipation while they have a lower intensity and seem to fade out more quickly on Fig.~\ref{fig:sla}.

\subsection{Libration}

\begin{figure}
\centering
\makebox[\textwidth][c]{\includegraphics[width=1.2\textwidth]{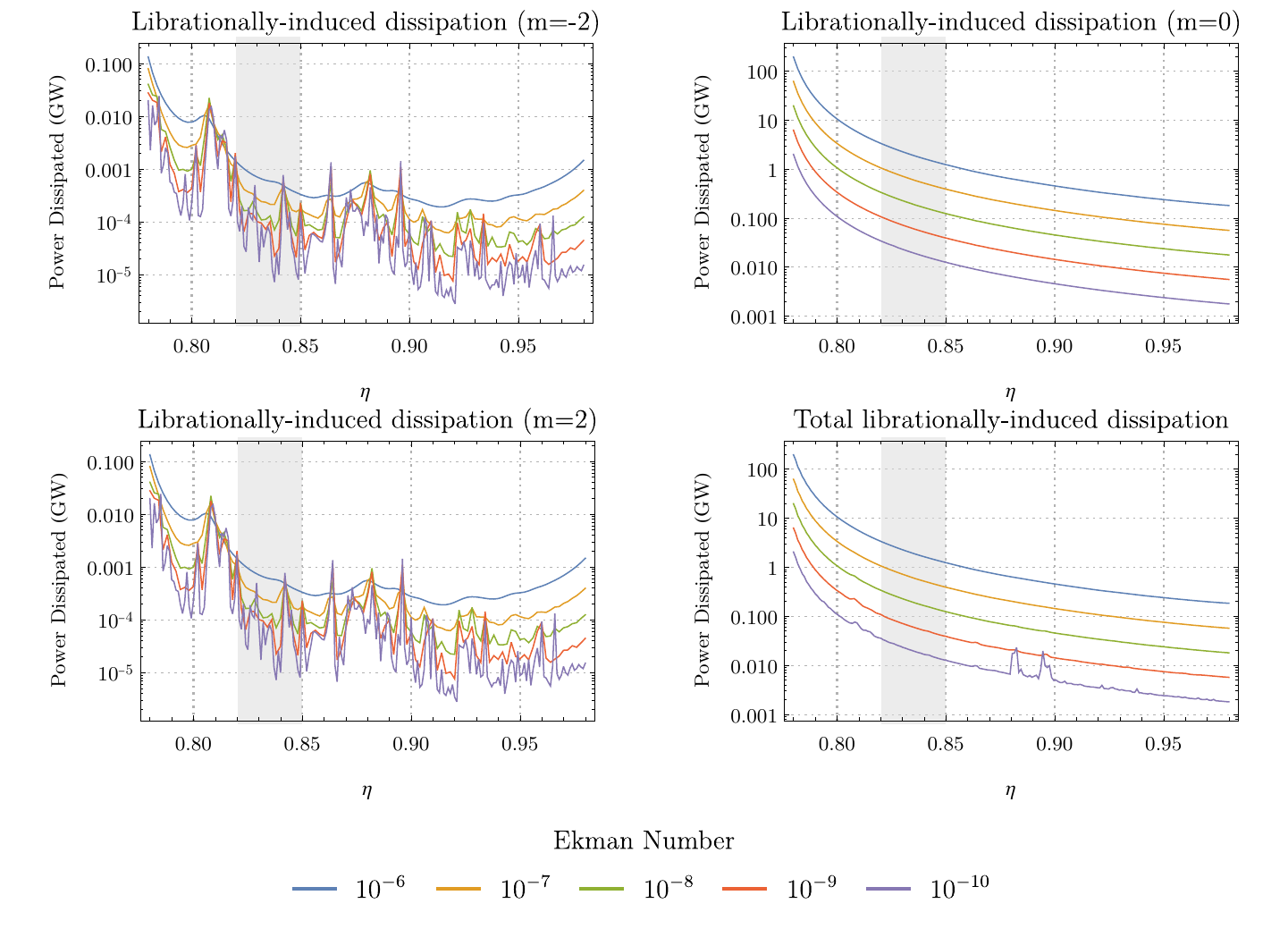}}%
\caption{{\bf Libration} -- Viscous dissipation in Enceladus's ocean as a function of $\eta$, the aspect ratio of the spherical shell, for different values of the Ekman number. The bottom-right plot is the sum of all the others. The shaded area represents the range $0.82\leq\eta\leq0.85$ predicted by models of isostasy.}
\label{fig:librationdissipation}
\end{figure}

Fig.~\ref{fig:librationdissipation} shows the dissipation for $m\in\{-2,0,2\}$ and the total dissipation for all these contributions (lower-right panel). We observe the same trend towards larger dissipation for decreasing $\eta$, similar to the situation with eccentricity tides. 

\begin{figure}
\centering
\includegraphics[width=0.6\textwidth]{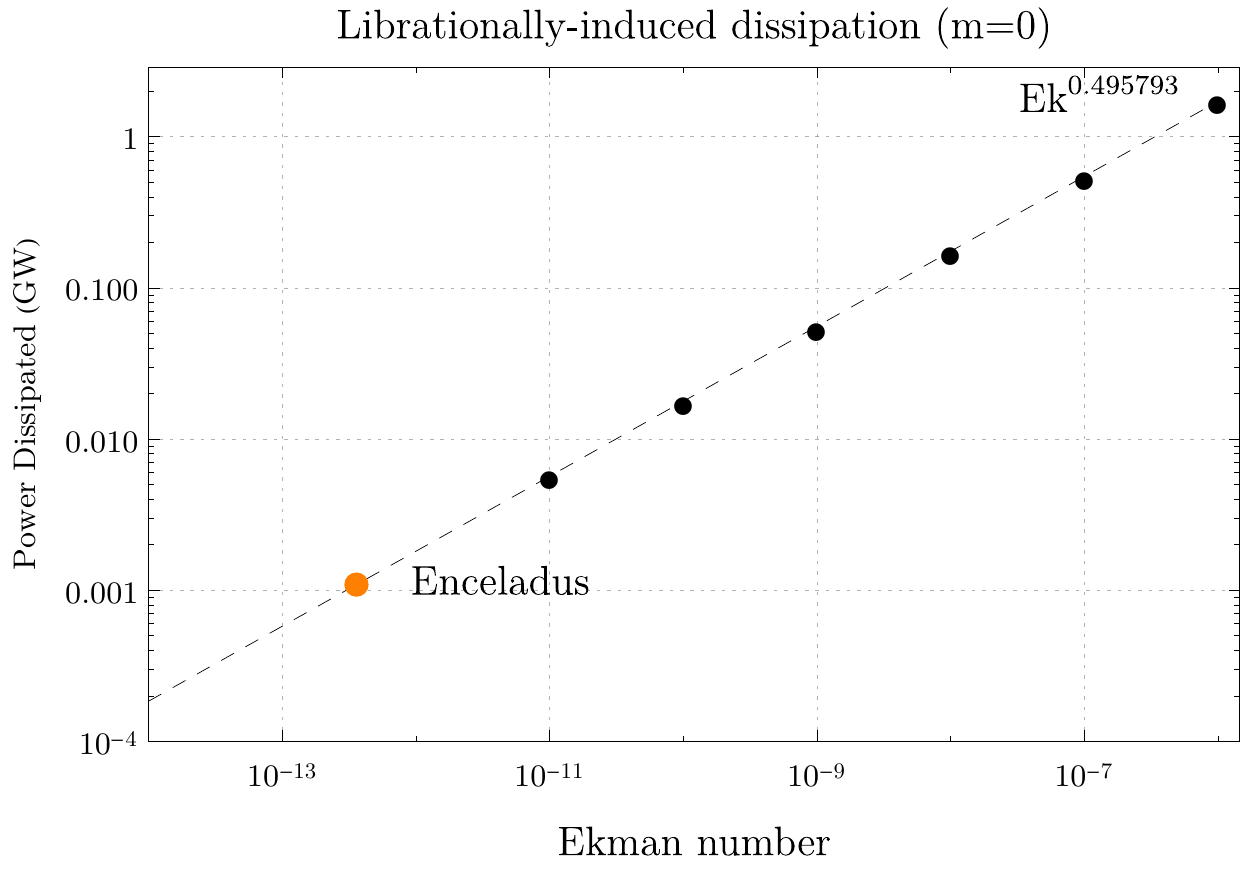}
\caption{{\bf Libration} -- Viscous dissipation in Enceladus's ocean ($\eta=0.838$). The behaviour $\sim \text{Ek}^{1/2}$ (dashed curve) indicates that most of the dissipation takes place inside the Ekman boundary layer by toroidal drag ($m=0$). This is not true of other components ($m=\pm2$, see Fig.~\ref{fig:librationdissipation}). The orange indicates the dissipation extrapolated to Enceladus's viscosity. This is the leading source of dissipation inside the moon's ocean.}
\label{fig:Dscaling_law_libm0}
\end{figure}

The contributions $m=2$ and $m=-2$ lead to identical dissipation profiles, as expected given the symmetrical nature of the forcing. The contribution $m=0$ dominates for larger values of the viscosity but decreases rapidly as $\text{Ek}$ goes down. This is the only contribution that shows no trace of peaks in its profile which indicates that most of the dissipation takes place inside the Ekman boundary layer. This picture is comforted by noting that the dissipation decreases with decreasing viscosity as $\mathcal{D}_{\text{visc}}\sim \text{Ek}^{1/2}$, as shown in Fig.~\ref{fig:Dscaling_law_libm0} for $\eta=0.838$, a scaling law typical of Ekman boundary layers. Extrapolation of this curve to $\text{Ek}=3.5\times10^{-13}$ gives $\mathcal{D}_\text{visc}\sim0.001~\text{GW}$ for Enceladus. This is the dominant source of dissipation in the moon's ocean if one disregards the other, more erratic, contributions. The total dissipation due to libration (Fig.~\ref{fig:librationdissipation}, lower-right pannel) also shows how the three components may become commensurable at low Ekman number ($\text{Ek}\leq10^{-10}$) at values of $\eta$ where there is a peak.

%% file: discussion.tex
\section{Discussion}
\label{sec:discussion}

\subsection{Libration as the dominant source of dissipation}

Toroidal libration ($m=0$) dominates over all other sources of dissipation for $\text{Ek}>10^{-10}$, as can be seen on Fig.~\ref{fig:tidaldissipation}~\&~\ref{fig:librationdissipation}, and becomes commensurable to the other components of libration for lower viscosities. Since the amount of dissipation induced by the toroidal drag scales nicely with the Ekman number, we can extrapolate the amount of dissipation to the Ekman number relevant to Enceladus from Fig.~\ref{fig:Dscaling_law_libm0}. The power scales as $\sim\text{Ek}^{1/2}$, which indicates the predominant role of the Ekman boundary layer. Extrapolating this power law to $\text{Ek}=3.5\times10^{-13}$, the relevant value for Enceladus, we predict a value of $10^{-3}$ GW for the total dissipation. That is about 4 orders of magnitude below the $\sim10$ GW observed by Cassini.

\subsection{The role of inertial modes}
\label{sec:inertialmodes}

Our results have revealed how the presence of internal shear layers can lead to a significant increase of the internal dissipation. Previous studies have attributed such peaks in the power profile to the existence of wave attractors \citep{Rieutord2000,Ogilvie2013,Rovira-Navarro2019}. We would like to complement this picture with our own interpretation based on the spectrum of free inertial modes.

In the absence of external body force, inertial modes are solutions to 
\begin{equation}
\lambda\vect{u}+2\vect{\hat{\Omega}}\times\vect{u}+\grad{p}-\text{Ek}\vect{\nabla}^2\vect{u}=0~,
\end{equation}
which is analogous to Eq.~(\ref{eq:NS}), except that $\lambda$ now denotes the \emph{complex} eigenvalue. It is sensible to assume that the response to a body force (per unit volume), $\vect{f}$, of frequency $\omega$ can be decomposed, at least partially, onto the (infinite) set of inertial modes $\{\vect{u}_\alpha\}$ :
\begin{equation}
\vect{v}\sim\sum_\alpha\frac{<\vect{f}|\vect{u}_\alpha>}{|\lambda_\alpha-i\omega|}~\vect{u}_\alpha~,
\end{equation}
where $<\cdot|\cdot>$ denotes the projection operator over pairs of vector fields. \citet{Ivers2014} and \citet{Backus2017} have demonstrated the completeness of the above modal expansion for an inviscid fluid inside a spherical or ellipsoidal container (with no inner core). In analogy with those works, we define the projection operator as $<\vect{v}|\vect{w}>\equiv \text{Re}\int_\mathcal{V}\vect{v}^*\cdot\vect{w}$. A resonance takes place when the factor multiplying one or more $\vect{u}_\alpha$ becomes large, which happens when the distance between the forcing frequency and the eigenvalue approaches zero and/or $<\vect{f}|\vect{u}>$ is large. This typically happens when the spatial structure of the forcing shares some similarity with an eigenmode. This formalism can be applied to our case after we have found the relevant body force, $\vect{f}$. Details on how to do so are explained in Appendix~\ref{sec:compRN2019}.
 
Fig.~\ref{fig:forced-modes} illustrates the resonance with inertial waves. The upper and middle panels represent the evolution of the spectrum around $\lambda=i$ as a function of $\eta$ ($\text{Ek}=10^{-7}$). The colour scale is inherited from the middle plot with lighter colours corresponding to a stronger damping. The lower panel shows a superposition of the dissipation profile (in black) on top of the combination $\frac{<\vect{f}|\vect{u}_\alpha>}{|\lambda_\alpha-i|}$. There is a clear correlation between the two; the peaks in dissipation profile appear where the resonances with the eigenmodes are the strongest. The trend towards more dissipation in the deep ocean limit can be attributed to a large amount of \emph{weak resonances} in this picture.

%
%

\begin{figure}
\centering
\includegraphics[width=\textwidth]{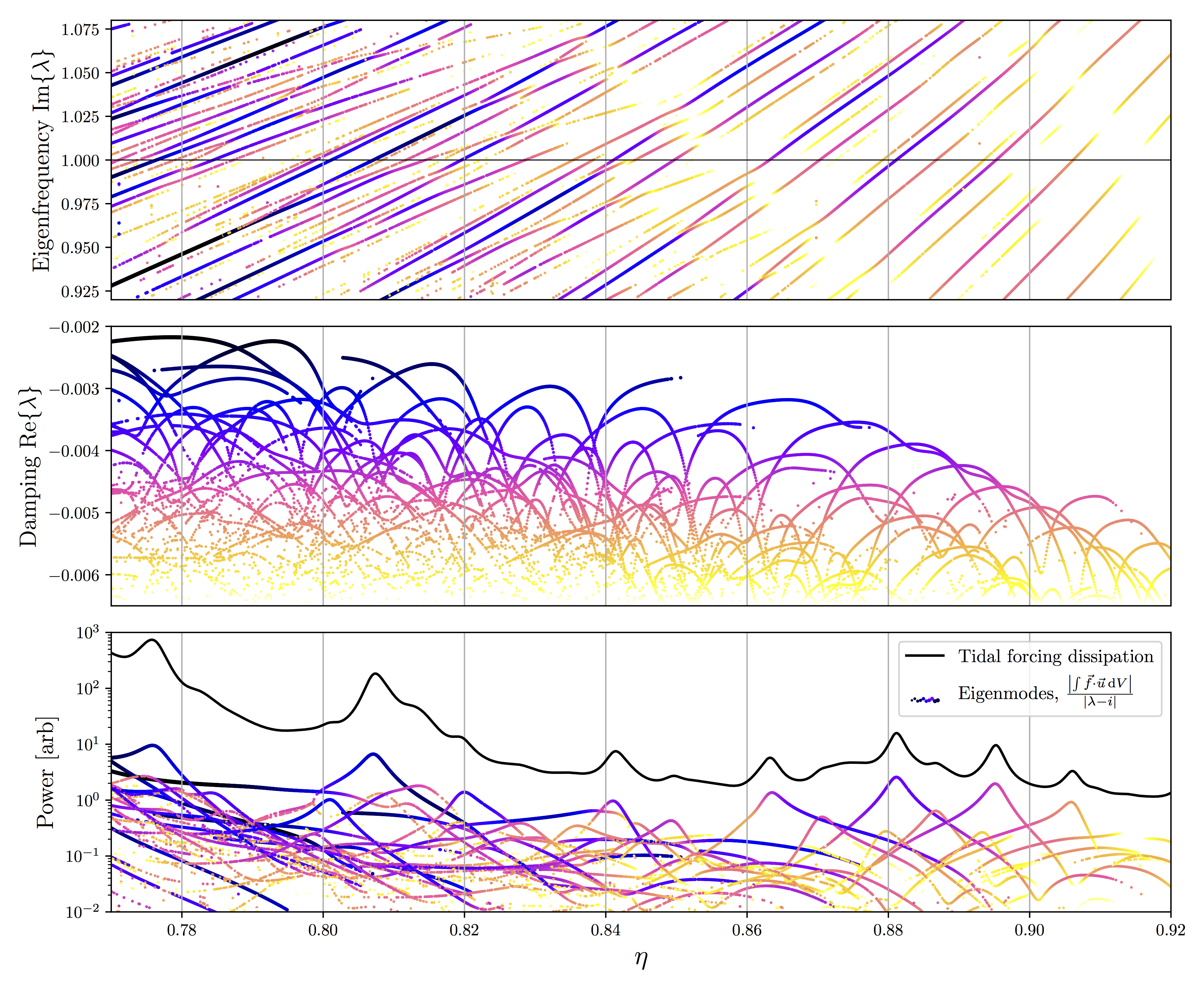}
\caption{{\bf Eccentricity tides} -- The upper and middle panels represent the imaginary part (frequency) and real part (damping) of the spectrum of eigenvalues of the linear Navier-Stokes equation respectively. The black curve of the lower panel represents the dissipated power (in arbitrary units). The colour curves on that same plot represent the projections of the tidal body force ($\vect{f}$) onto the eigenvectors ($\vect{u}$) divided by the distance between the unit forcing frequency and the eigenvalue ($\lambda$). The colour scale is inherited from the middle panel with lighter colours corresponding to a stronger damping ($m=2, \text{Ek}=10^{-7}$).}
\label{fig:forced-modes}
\end{figure}

\subsection{Conclusion and Future work}

We have computed the viscous dissipation in Enceladus's ocean and we have shown that the combined effect of eccentricity tides and libration is not sufficient to explain the heat flux observed by Cassini. We therefore confirm that the results of \citet{Matsuyama2018} remain valid even when the dissipation of the internal shear layers in the ocean's bulk are taken into account. We show that librationally-induced dissipation dominates tidally-induced dissipation, but the overall power dissipated in the ocean is still negligible, as in \citet{Rovira-Navarro2019}. 

In light of our results, it is clear that the heat flux observed on Enceladus cannot be explained solely by viscous dissipation in the ocean alone, at least at the linear level. The source of heat must therefore be sought elsewhere. The ohmic dissipation in the shear layers due to the presence of Saturn's magnetic field appears as a relevant candidate at first glance. However, the external magnetic field at Enceladus is quite small, of the order $\sim 0.6~\mu$T, which corresponds to a \emph{Lehnert number} of $\text{Le}\sim1.6\times10^{-6}$. \citet{Lin2018} have shown that ohmic dissipation dominates over viscous dissipation only when $\text{Le}>\text{Ek}_m^{2/3}$, where $\text{Ek}_m$ is the \emph{magnetic Ekman number}\footnote{The Lehnert and magnetic Ekman numbers are defined as \begin{align*}\text{Le}=\frac{B_0}{\Omega R_o \sqrt{\mu \rho}}&&\text{Ek}_m=\frac{\nu_m}{\Omega R_o^2}\end{align*} where $B_0$ is the background magnetic field, $\mu$ is the magnetic permeability and $\nu_m$ is the magnetic diffusivity of the fluid.}. Recent estimates of the electrical conductivity inside the ocean \citep{Vance2018} give $\text{Ek}_m\sim1$. The ohmic dissipation is therefore unlikely to contribute significantly to the total.

\citet{Wilson2018} have argued that libration could potentially generate turbulence inside the ocean and thus significantly increase the amount of dissipation.

Another possibility is that the present heat production does not balance the present heat flow, but was larger in the past during periods of larger eccentricities \citep{Ojakangas1986}.

The observed North-South dichotomy of the moon's surface poses another important puzzle. The southern hemisphere is young and re-surfaced, while the northern hemisphere is old and cratered. This problem cannot be addressed in our simplified spherical model. The observed $J_3$ implies that the icy crust is thinner over the south polar terrain \citep{Iess2014}. We also know that the density of kinetic energy inside the shear layers is higher at the poles, close to the axis of rotation \citep{Rieutord1997}, something that is visible on Fig. \ref{fig:shearlayers}. It would be interesting to see how the dissipation at both poles changes when one takes the topography of the ocean into account. On the one hand, the internal shear layers tend to be more focused in the thin ocean limit. On the other hand, the overall dissipated power tends to be higher in the thick ocean limit. Both regimes are illustrated on the left and right sides of every dissipation profile in the present paper (e.g. Fig.~\ref{fig:tidaldissipation}~\&~\ref{fig:librationdissipation}).


%% file: acknowledgement.tex
\section*{Acknowledgement}
The research leading to these results has received funding from the European Research Council (ERC) under the European Union's Horizon 2020 research and innovation programme (Advanced Grant agreement No~670874).

%% file: appendix.tex
\begin{appendix}

\section{Tidal and libration forcings}
\subsection{Tidal displacements}
\label{sec:tidaldisplacements}

\begin{figure}
\centering
\begin{tabular}{cc}
\subfloat{\includegraphics[width=0.45\textwidth]{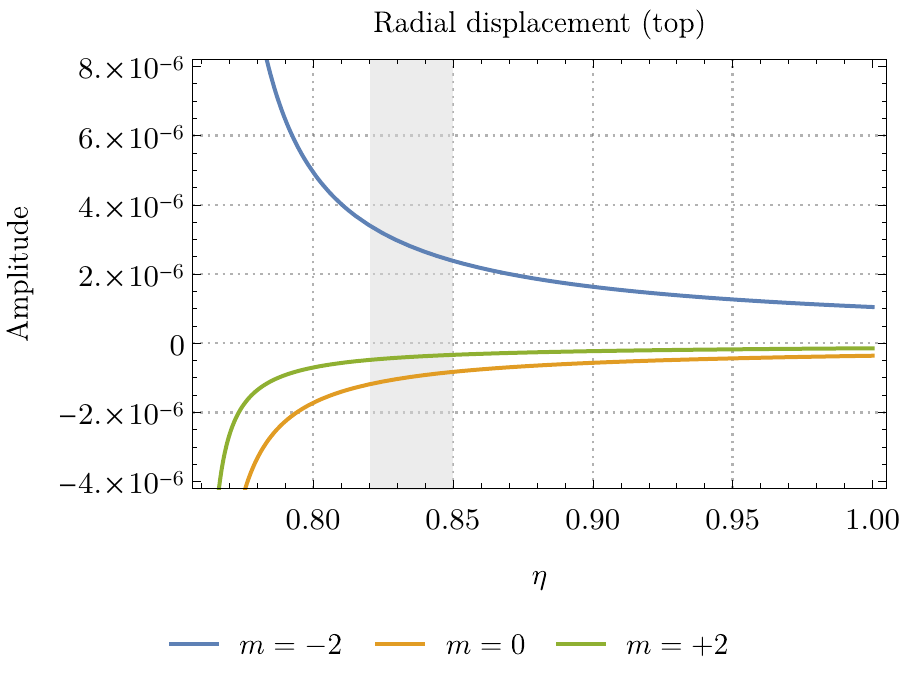}} &
\subfloat{\includegraphics[width=0.45\textwidth]{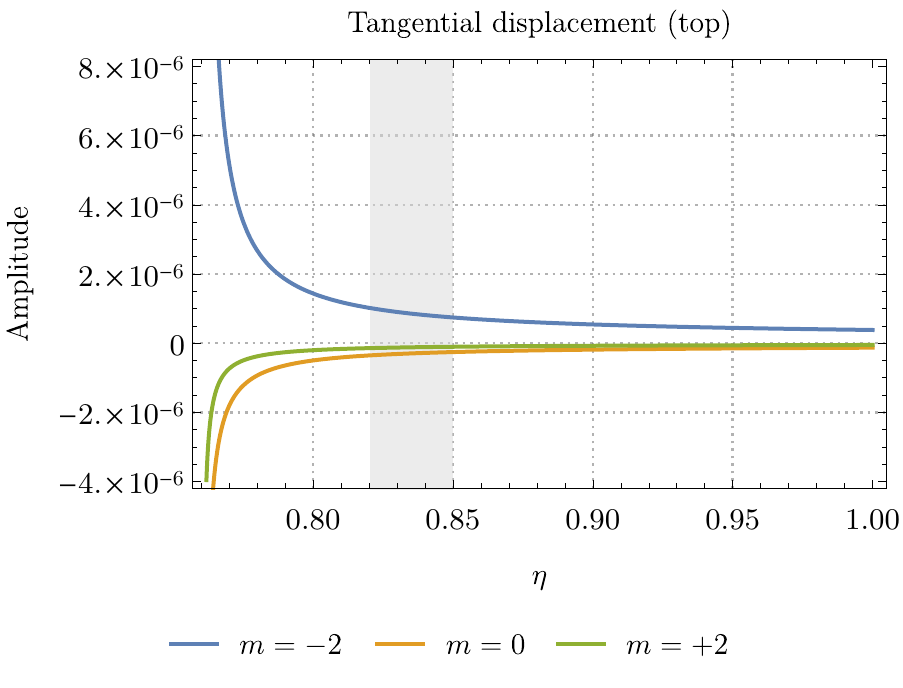}}\\
\subfloat{\includegraphics[width=0.45\textwidth]{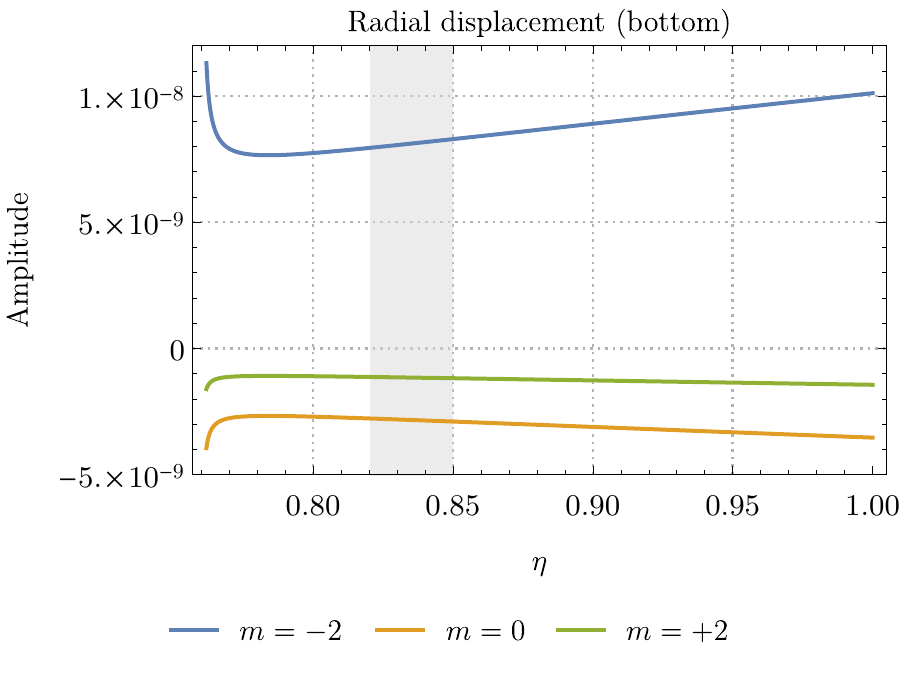}} &
\subfloat{\includegraphics[width=0.45\textwidth]{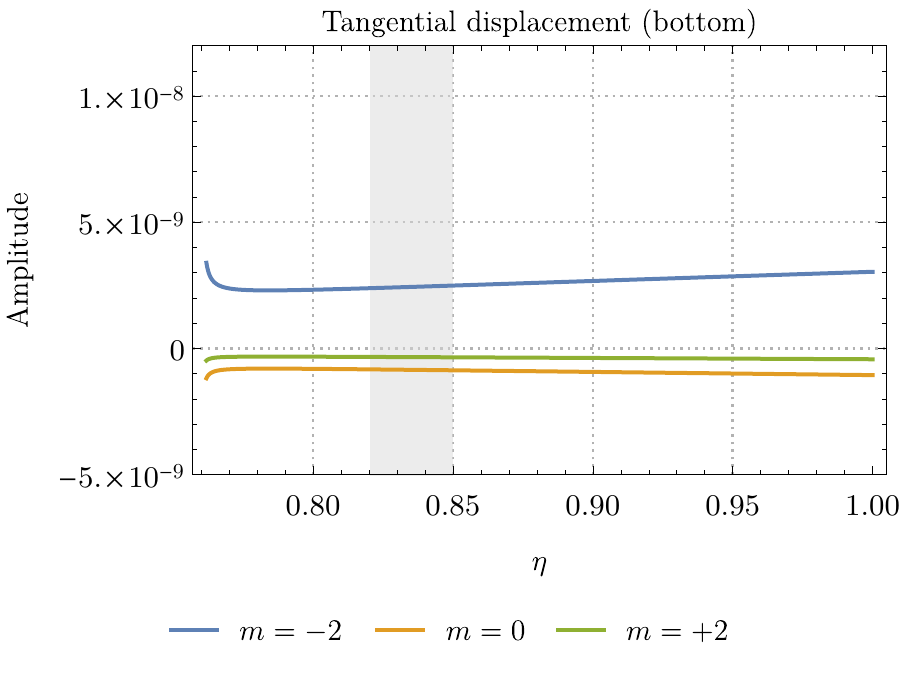}}\\
\end{tabular}
\caption{Tidal displacements at the top and bottom of the ocean as a function of the aspect ratio $\eta$. The values of the amplitudes are given in units of outer radius of the ocean (base of the icy shell). The shaded area represents the range $0.82\leq\eta\leq0.85$ predicted by models of isostasy.}
\label{fig:tidaldisplacements}
\end{figure}

The three spherical harmonics components of the radial and tangential displacements are shown on Fig.~\ref{fig:tidaldisplacements} for different values of the aspect ratio, $\eta$. 
The analytical expressions for the displacement are quite long and impractical. Instead, we provide a set of simplified expressions obtained using a Pade approximant on the interval $\eta\in[\frac{16}{21},1]$. All expressions are in dimensionless units~:
\vspace{0.2in}

\noindent
\underline{radial displacement at the top of the ocean}
\scriptsize
\begin{align}
m=-2~\rightarrow&&\frac{\left(1.24689\times 10^{-5}\right) \left(\eta
   -\frac{37}{42}\right)^3+\left(1.20137\times 10^{-5}\right) \left(\eta
   -\frac{37}{42}\right)^2+\left(5.9538\times 10^{-6}\right) \left(\eta
   -\frac{37}{42}\right)+1.80557\times 10^{-6}}{49.5224 \left(\eta
   -\frac{37}{42}\right)^3+22.7936 \left(\eta -\frac{37}{42}\right)^2+10.1381 \left(\eta
   -\frac{37}{42}\right)+0.979681}~,\\
m=0~\rightarrow&&\frac{\left(-4.36321\times 10^{-6}\right) \left(\eta
   -\frac{37}{42}\right)^3-\left(4.20393\times 10^{-6}\right) \left(\eta
   -\frac{37}{42}\right)^2-\left(2.0834\times 10^{-6}\right) \left(\eta
   -\frac{37}{42}\right)-6.31816\times 10^{-7}}{\left(4.95224\times 10^1\right)
   \left(\eta -\frac{37}{42}\right)^3+\left(2.27936\times 10^1\right) \left(\eta
   -\frac{37}{42}\right)^2+\left(1.01381\times 10^1\right) \left(\eta
   -\frac{37}{42}\right)+9.79681\times 10^{-1}}~,\\
m=2~\rightarrow&&\frac{\left(-1.78127\times 10^{-6}\right) \left(\eta
   -\frac{37}{42}\right)^3-\left(1.71625\times 10^{-6}\right) \left(\eta
   -\frac{37}{42}\right)^2-\left(8.50543\times 10^{-7}\right) \left(\eta
   -\frac{37}{42}\right)-2.57938\times 10^{-7}}{\left(4.95224\times 10^1\right)
   \left(\eta -\frac{37}{42}\right)^3+\left(2.27936\times 10^1\right) \left(\eta
   -\frac{37}{42}\right)^2+\left(1.01381\times 10^1\right) \left(\eta
   -\frac{37}{42}\right)+9.79681\times 10^{-1}}~.
\end{align}
\normalsize
\underline{radial displacement at the bottom of the ocean}
\scriptsize
\begin{align}
m=-2~\rightarrow&&\frac{\left(1.29729\times 10^{-7}\right) \left(\eta
   -\frac{37}{42}\right)^3+\left(2.28975\times 10^{-7}\right) \left(\eta
   -\frac{37}{42}\right)^2+\left(9.73595\times 10^{-8}\right) \left(\eta
   -\frac{37}{42}\right)+8.70387\times 10^{-9}}{\left(-1.33268\times 10^{-1}\right)
   \left(\eta -\frac{37}{42}\right)^3+\left(1.24677\times 10^1\right) \left(\eta
   -\frac{37}{42}\right)^2+9.81601 \left(\eta -\frac{37}{42}\right)+1.00405}~,\\
m=0~\rightarrow&&\frac{\left(-4.53957\times 10^{-8}\right) \left(\eta
   -\frac{37}{42}\right)^3-\left(8.01246\times 10^{-8}\right) \left(\eta
   -\frac{37}{42}\right)^2-\left(3.40687\times 10^{-8}\right) \left(\eta
   -\frac{37}{42}\right)-3.04572\times 10^{-9}}{\left(-1.33268\times 10^{-1}\right)
   \left(\eta -\frac{37}{42}\right)^3+\left(1.24677\times 10^1\right) \left(\eta
   -\frac{37}{42}\right)^2+9.81601 \left(\eta -\frac{37}{42}\right)+1.00405}~,\\
m=2~\rightarrow&&\frac{\left(-1.85327\times 10^{-8}\right) \left(\eta
   -\frac{37}{42}\right)^3-\left(3.27107\times 10^{-8}\right) \left(\eta
   -\frac{37}{42}\right)^2-\left(1.39085\times 10^{-8}\right) \left(\eta
   -\frac{37}{42}\right)-1.24341\times 10^{-9}}{\left(-1.33268\times 10^{-1}\right)
   \left(\eta -\frac{37}{42}\right)^3+\left(1.24677\times 10^1\right) \left(\eta
   -\frac{37}{42}\right)^2+9.81601 \left(\eta -\frac{37}{42}\right)+1.00405}~.
\end{align}
\normalsize
\underline{tangential displacement at the top of the ocean}
\scriptsize
\begin{align}
m=-2~\rightarrow&&\frac{\left(7.45962\times 10^{-6}\right) \left(\eta
   -\frac{37}{42}\right)^3+\left(6.64185\times 10^{-6}\right) \left(\eta
   -\frac{37}{42}\right)^2+\left(2.78583\times 10^{-6}\right) \left(\eta
   -\frac{37}{42}\right)+5.73503\times 10^{-7}}{\left(5.90717\times 10^1\right)
   \left(\eta -\frac{37}{42}\right)^3+\left(2.65663\times 10^1\right) \left(\eta
   -\frac{37}{42}\right)^2+\left(1.03018\times 10^1\right) \left(\eta
   -\frac{37}{42}\right)+9.61236\times 10^{-1}}~,\\
m=0~\rightarrow&&\frac{\left(-2.61032\times 10^{-6}\right) \left(\eta
   -\frac{37}{42}\right)^3-\left(2.32416\times 10^{-6}\right) \left(\eta
   -\frac{37}{42}\right)^2-\left(9.74839\times 10^{-7}\right) \left(\eta
   -\frac{37}{42}\right)-2.00684\times 10^{-7}}{\left(5.90717\times 10^1\right)
   \left(\eta -\frac{37}{42}\right)^3+\left(2.65663\times 10^1\right) \left(\eta
   -\frac{37}{42}\right)^2+\left(1.03018\times 10^1\right) \left(\eta
   -\frac{37}{42}\right)+9.61236\times 10^{-1}}~,\\
m=2~\rightarrow&&\frac{\left(-1.06566\times 10^{-6}\right) \left(\eta
   -\frac{37}{42}\right)^3-\left(9.48836\times 10^{-7}\right) \left(\eta
   -\frac{37}{42}\right)^2-\left(3.97976\times 10^{-7}\right) \left(\eta
   -\frac{37}{42}\right)-8.19289\times 10^{-8}}{\left(5.90717\times 10^1\right)
   \left(\eta -\frac{37}{42}\right)^3+\left(2.65663\times 10^1\right) \left(\eta
   -\frac{37}{42}\right)^2+\left(1.03018\times 10^1\right) \left(\eta
   -\frac{37}{42}\right)+9.61236\times 10^{-1}}~.
\end{align}
\normalsize
\underline{tangential displacement at the bottom of the ocean}
\scriptsize
\begin{align}
m=-2~\rightarrow&&\frac{\left(3.89187\times 10^{-8}\right) \left(\eta
   -\frac{37}{42}\right)^3+\left(6.86926\times 10^{-8}\right) \left(\eta
   -\frac{37}{42}\right)^2+\left(2.92078\times 10^{-8}\right) \left(\eta
   -\frac{37}{42}\right)+2.61116\times 10^{-9}}{\left(-1.33268\times 10^{-1}\right)
   \left(\eta -\frac{37}{42}\right)^3+\left(1.24677\times 10^1\right) \left(\eta
   -\frac{37}{42}\right)^2+9.81601 \left(\eta -\frac{37}{42}\right)+1.00405}~,\\
m=0~\rightarrow&&\frac{\left(-1.36187\times 10^{-8}\right) \left(\eta
   -\frac{37}{42}\right)^3-\left(2.40374\times 10^{-8}\right) \left(\eta
   -\frac{37}{42}\right)^2-\left(1.02206\times 10^{-8}\right) \left(\eta
   -\frac{37}{42}\right)-9.13716\times 10^{-10}}{\left(-1.33268\times 10^{-1}\right)
   \left(\eta -\frac{37}{42}\right)^3+\left(1.24677\times 10^1\right) \left(\eta
   -\frac{37}{42}\right)^2+9.81601 \left(\eta -\frac{37}{42}\right)+1.00405}~,\\
m=2~\rightarrow&&\frac{\left(-5.55981\times 10^{-9}\right) \left(\eta
   -\frac{37}{42}\right)^3-\left(9.81322\times 10^{-9}\right) \left(\eta
   -\frac{37}{42}\right)^2-\left(4.17255\times 10^{-9}\right) \left(\eta
   -\frac{37}{42}\right)-3.73023\times 10^{-10}}{\left(-1.33268\times 10^{-1}\right)
   \left(\eta -\frac{37}{42}\right)^3+\left(1.24677\times 10^1\right) \left(\eta
   -\frac{37}{42}\right)^2+9.81601 \left(\eta -\frac{37}{42}\right)+1.00405}~.
\end{align}
\normalsize

\subsection{Libration forcing}
\label{sec:librationforcing}
\begin{figure}
\centering
\begin{tabular}{ccc}
\subfloat{\includegraphics[width=0.48\textwidth]{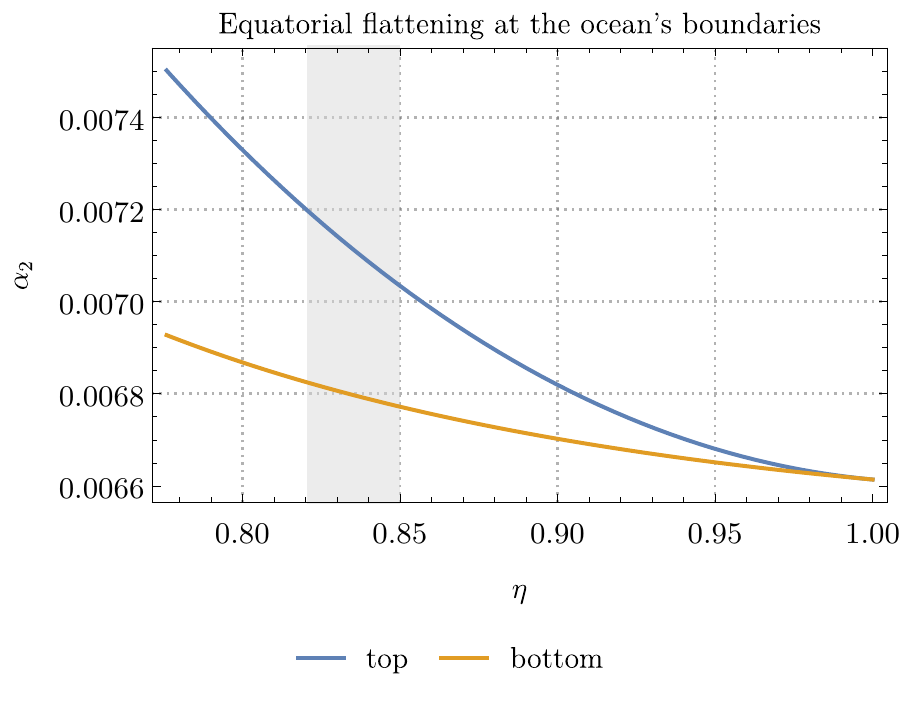}\label{fig:triaxial_flattening}}&
\subfloat{\includegraphics[width=0.48\textwidth]{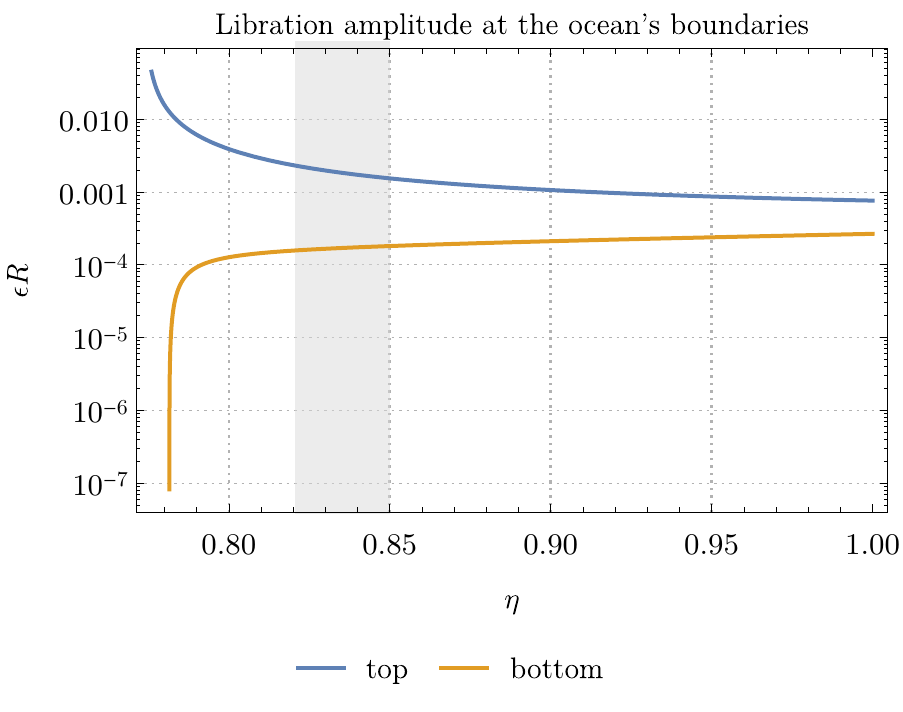}\label{fig:libration_amplitude}}
\end{tabular}
\caption{Equatorial flattening (left panel) and amplitude of libration (right panel) at the top and bottom of the ocean. The shaded area represents the range $0.82\leq\eta\leq0.85$ predicted by models of isostasy.}
\label{fig:libration_parameters}
\end{figure}

The value of the parameters $\alpha_2$ and $\epsilon R$ at the top and bottom of the ocean are represented on Fig.~\ref{fig:libration_parameters}. 
The analytical expressions for the amplitude $\epsilon R$ are quite long and impractical. And so we provide an approximation on the interval $\eta\in[\frac{16}{21},1]$ based on a Pade approximant (all expressions are in dimensionless units)~:\\

\noindent\underline{Amplitude at the top and bottom of the ocean}
\scriptsize
\begin{align}
\text{top}~\rightarrow&&\epsilon R &=\frac{0.0142482 \left(\eta -\frac{37}{42}\right)^4+0.0234122 \left(\eta -\frac{37}{42}\right)^3+0.0177355 \left(\eta
   -\frac{37}{42}\right)^2+0.00669395 \left(\eta -\frac{37}{42}\right)+0.00113254}{25.2386 \left(\eta -\frac{37}{42}\right)^4+45.0416 \left(\eta
   -\frac{37}{42}\right)^3+33.2167 \left(\eta -\frac{37}{42}\right)^2+12.0145 \left(\eta -\frac{37}{42}\right)+0.957481}\\
\text{bottom}~\rightarrow&&\epsilon R &=\frac{0.00177273 \left(\eta -\frac{37}{42}\right)^4+0.00779929 \left(\eta -\frac{37}{42}\right)^3+0.00780726 \left(\eta
   -\frac{37}{42}\right)^2+0.00272733 \left(\eta -\frac{37}{42}\right)+0.000201493}{0.997077 \left(\eta -\frac{37}{42}\right)^4-4.0871 \left(\eta
   -\frac{37}{42}\right)^3+11.8146 \left(\eta -\frac{37}{42}\right)^2+10.7977 \left(\eta -\frac{37}{42}\right)+1.01542}~.
\end{align}
\normalsize

\section{Tidal dissipation with a body force}
\label{sec:compRN2019} 

Some authors \citep{Rieutord2000,Ogilvie2013,Rovira-Navarro2019} presented a slightly different formalism to compute the dissipation caused by tides. We now want to show the equivalence to ours. Their approach is based on the decomposition of the velocity field in two parts :
\begin{equation}
\vect{v}=\vect{v}^{\text{(eq)}}+\vect{v}^{\text{(d)}}~.
\end{equation}
The first part is attributed to the displacement caused by the equilibrium tides. Plugging the above ansatz into the momentum equation, leads to an equation for $\vect{v}^\text{(d)}$~:
\begin{equation}
i\omega\vect{v}^\text{(d)}+2\vect{\hat{\Omega}}\times\vect{v}^\text{(d)}+\grad{p}-\text{Ek}\nabla^2\vect{v}^\text{(d)}=\vect{f}~,
\end{equation}
where $\vect{f}=-i\omega\vect{v}^\text{(eq)}-2\vect{\hat{\Omega}}\times\vect{v}^\text{(eq)}$ is treated as a body force per unit volume. \citet{Ogilvie2013} argues that when $\vect{v}^\text{(eq)}$ is computed from the theory of equilibrium tides, one has $\vect{v}^\text{(eq)}=\grad{X}$, where $X$ is an harmonic scalar function ($\vect{\nabla}^2X=0$, owing to the condition of incompressibility). The harmonic component of degree $\ell$ therefore reads $X_{\ell,m}=(A_{\ell,m} r^\ell+\frac{B_{\ell,m}}{r^{\ell+1}})$, where $A_{\ell,m}$ and $B_{\ell,m}$ are scalar constants that depend on the deformation (see \citet{Ogilvie2013} for details).

In our approach, we do not decompose the velocity field explicitly and rather solve for the full momentum equation, including the potential perturbation. This extra term, however, can be merged with the reduced pressure, which, in turn, disappears completely when one solves for the vorticity in the volume. The effect of the potential thus appears only in the expression of the boundary condition. This is the reason for which, the condition of continuity of the total flow, $\vect{v}$, across the physical boundaries is given by our Eq.~(\ref{eq:bcv}) while \citet{Rovira-Navarro2019} simply have $\vect{v}^{(d)}=0$.

The two methods lead to results that are completely equivalent. The approach based on the boundary forcing is perhaps conceptually simpler and better suited to the usual language of fluid dynamicists. The description based on the body force $\vect{f}$, on the other hand, has the advantage to be directly usable in the discussion presented in Sec.~\ref{sec:inertialmodes}.

\begin{figure}
\centering
\includegraphics[width=0.7\textwidth]{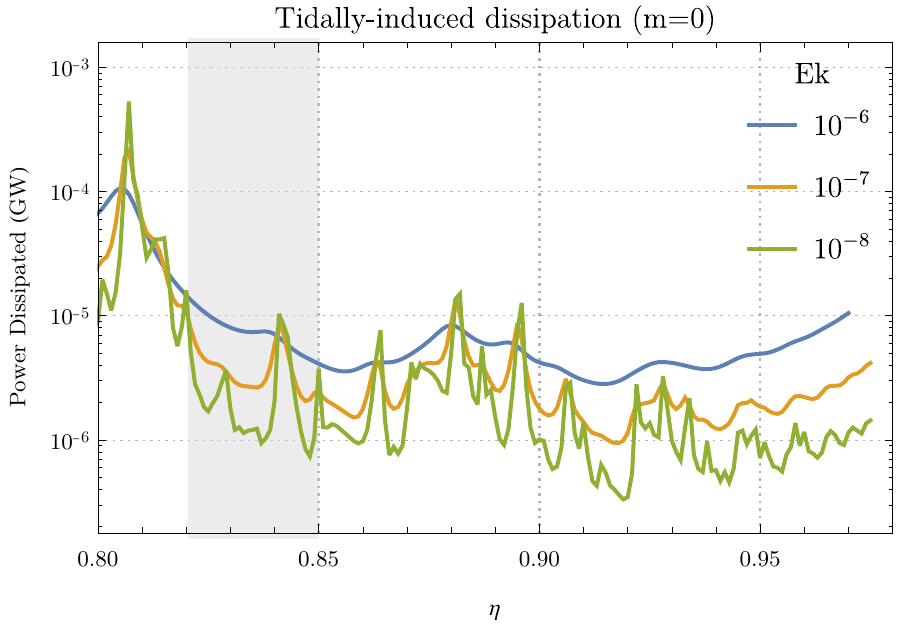}
\caption{{\bf Eccentricity tides} -- Detailed version of the upper-right panel of Fig.\ref{fig:tidaldissipation} for comparison with \citet{Rovira-Navarro2019} (see their Fig.6 (b)).}
\label{fig:tide_D_RoviraNavarro}
\end{figure}

Fig.~\ref{fig:tide_D_RoviraNavarro} is a reproduction of Fig.6 (b) in \citet{Rovira-Navarro2019} using our settings. The discrepancy between the predicted amounts of dissipation in the deep ocean limit (left part of our plot, right part of theirs) is due to the presence of an icy crust in our model while it is neglected in their study. This leads to a decrease in the amount of energy dissipated in agreement with the results of \citet{Matsuyama2018}. Our curves also go up slightly in the thin ocean limit on the right part of the plot. This feature is absent in \citet{Rovira-Navarro2019} and is likely due to the fact that they use a purely radial forcing, thus neglecting tangential displacement at the boundary. When we take these into account, resonances appear in the thin ocean limit consistently with the results of \citet{Ogilvie2009}. Appart from these differences, our curves bear a striking ressemblance with theirs, which illustrates the equivalence of our approaches.

\section{Navier-Stokes equation in spherical harmonics}
\label{sec:NSYlm}
Here below, we provide the expressions used to compute the flow inside the ocean. These are based on the curl of Eq.~(\ref{eq:NS}), \emph{i.e.} the equation of vorticity, and the ansatz  Eq.~(\ref{eq:poltor})~:
\footnotesize
\begin{align}
r^4(\hat{\vect{r}}\cdot\vect{\nabla}\times\vect{\nabla}\times\text{Eq.}(\ref{eq:NS})) = 
&~ir^2(\omega\ell(\ell+1)-2m)\left[\ell(\ell+1)P_{\ell,m}-2rP_{\ell,m}'-r^2P_{\ell,m}''\right]\nonumber\\
&+\frac{2 r^3 (\ell -1)^2 (\ell +1) \sqrt{(\ell -\mathit{m}) (\mathit{m}+\ell )}}{2 \ell-1}~T_{\ell-1,m}-\frac{2 r^3 \ell  (\ell +2)^2 \sqrt{(-\mathit{m}+\ell +1) (\mathit{m}+\ell +1)}}{2 \ell+3}~T_{\ell+1,m}\nonumber\\
&-\frac{2 r^4 \left(\ell ^2-1\right) \sqrt{(\ell -\mathit{m}) (\mathit{m}+\ell )}}{2 \ell -1}~T_{\ell-1,m}'-\frac{2 r^4 \ell  (\ell +2) \sqrt{(-\mathit{m}+\ell +1) (\mathit{m}+\ell +1)}}{2 \ell +3}~T_{\ell+1,m}'\nonumber\\
&+\text{Ek}~\ell(\ell+1)\left[(\ell-1)\ell(\ell+1)(\ell+2)P_{\ell,m}-2r^2\ell(\ell+1)P_{\ell,m}''+4r^3P_{\ell,m}'''+r^4P_{\ell,m}''''\right]\nonumber\\=&~0~,\label{eq:curlcurlNS}\\
-r^2(\hat{\vect{r}}\cdot\vect{\nabla}\times\text{Eq.}(\ref{eq:NS})) = 
&~ir^2( \omega  \ell  (\ell +1)-2 \mathit{m})~T_{\ell ,m}\nonumber\\
&+\frac{2 r (\ell -1)^2 (\ell +1) \sqrt{(\ell -\mathit{m}) (\mathit{m}+\ell )}}{2 \ell -1}~P_{\ell-1,m}-\frac{2 r \ell  (\ell +2)^2 \sqrt{(-\mathit{m}+\ell +1) (\mathit{m}+\ell +1)}}{2 \ell +3}~P_{\ell+1,m}\nonumber\\
&-\frac{2 r^2 \left(\ell ^2-1\right) \sqrt{(\ell -\mathit{m}) (\mathit{m}+\ell )}}{2 \ell -1}~P_{\ell-1,m}'-\frac{2 r^2 \ell  (\ell +2) \sqrt{(-\mathit{m}+\ell +1) (\mathit{m}+\ell +1)}}{2 \ell +3}~P_{\ell+1,m}'\nonumber\\
&+\text{Ek}~\ell  (\ell +1) \left[\ell  (\ell +1) T_{\ell ,m}-2r
   T_{\ell ,m}'-r^2 T_{\ell ,m}''\right]\nonumber\\=&~0~.\label{eq:curlNS}
\end{align}
\normalsize

\end{appendix}

%% file: main.bbl
\begin{thebibliography}{37}
\providecommand{\natexlab}[1]{#1}
\providecommand{\url}[1]{\texttt{#1}}
\expandafter\ifx\csname urlstyle\endcsname\relax
  \providecommand{\doi}[1]{doi: #1}\else
  \providecommand{\doi}{doi: \begingroup \urlstyle{rm}\Url}\fi

\bibitem[Backus and Rieutord(2017)]{Backus2017}
G.~Backus and M.~Rieutord.
\newblock {Completeness of inertial modes of an incompressible inviscid fluid
  in a corotating ellipsoid}.
\newblock \emph{Physical Review E}, 95\penalty0 (5):\penalty0 1--16, 2017.

\bibitem[Baland and {Van Hoolst}(2010)]{Baland2010}
R.~M. Baland and T.~{Van Hoolst}.
\newblock {Librations of the Galilean satellites: The influence of global
  internal liquid layers}.
\newblock \emph{Icarus}, 209\penalty0 (2):\penalty0 651--664, 2010.

\bibitem[Baland et~al.(2016)Baland, Yseboodt, and {Van Hoolst}]{Baland2016}
R.~M. Baland, M.~Yseboodt, and T.~{Van Hoolst}.
\newblock {The obliquity of Enceladus}.
\newblock \emph{Icarus}, 268:\penalty0 12--31, 2016.

\bibitem[Barr and McKinnon(2007)]{Barr2007}
A.~C. Barr and W.~B. McKinnon.
\newblock {Convection in Enceladus' ice shell: Conditions for initiation}.
\newblock \emph{Geophysical Research Letters}, 34\penalty0 (9):\penalty0 2--7,
  2007.

\bibitem[B{\v{e}}hounkov{\'{a}} et~al.(2010)B{\v{e}}hounkov{\'{a}}, Tobie,
  Choblet, and {\v{C}}adek]{Behounkova2010}
M.~B{\v{e}}hounkov{\'{a}}, G.~Tobie, G.~Choblet, and O.~{\v{C}}adek.
\newblock {Coupling mantle convection and tidal dissipation: Applications to
  Enceladus and Earth-like planets}.
\newblock \emph{Journal of Geophysical Research E: Planets}, 115\penalty0
  (9):\penalty0 1--20, 2010.

\bibitem[B{\v{e}}hounkov{\'{a}} et~al.(2017)B{\v{e}}hounkov{\'{a}},
  Sou{\v{c}}ek, Hron, and {\v{C}}adek]{Behounkova2017}
M.~B{\v{e}}hounkov{\'{a}}, O.~Sou{\v{c}}ek, J.~Hron, and O.~{\v{C}}adek.
\newblock {Plume Activity and Tidal Deformation on Enceladus Influenced by
  Faults and Variable Ice Shell Thickness}.
\newblock \emph{Astrobiology}, 17\penalty0 (9):\penalty0 941--954, 2017.

\bibitem[Beuthe et~al.(2016)Beuthe, Rivoldini, and Trinh]{Beuthe2016}
M.~Beuthe, A.~Rivoldini, and A.~Trinh.
\newblock {Enceladus's and Dione's floating ice shells supported by minimum
  stress isostasy}.
\newblock \emph{Geophysical Research Letters}, 43\penalty0 (19):\penalty0
  10,088--10,096, 2016.

\bibitem[Chen and Nimmo(2011)]{Chen2011}
E.~M.~A. Chen and F.~Nimmo.
\newblock {Obliquity tides do not significantly heat Enceladus}.
\newblock \emph{Icarus}, 214\penalty0 (2):\penalty0 779--781, 2011.

\bibitem[Choblet et~al.(2017)Choblet, Tobie, Sotin, B{\v{e}}hounkov{\'{a}},
  {\v{C}}adek, Postberg, and Sou{\v{c}}ek]{Choblet2017}
G.~Choblet, G.~Tobie, C.~Sotin, M.~B{\v{e}}hounkov{\'{a}}, O.~{\v{C}}adek,
  F.~Postberg, and O.~Sou{\v{c}}ek.
\newblock {Powering prolonged hydrothermal activity inside Enceladus}.
\newblock \emph{Nature Astronomy}, 1\penalty0 (12):\penalty0 841--847, 2017.

\bibitem[Dahlen and Tromp(1998)]{dahlen1998}
F.~Dahlen and J.~Tromp.
\newblock \emph{Theoretical Global Seismology}.
\newblock Princeton University Press, 1998.

\bibitem[Greenspan(1968)]{greenspan1968}
H.~P. Greenspan.
\newblock \emph{The Theory of Rotating Fluids}.
\newblock Cambridge Monographs on Mechanics. Cambridge University Press, 1968.

\bibitem[Hay and Matsuyama(2017)]{Hay2017}
H.~C. Hay and I.~Matsuyama.
\newblock {Numerically modelling tidal dissipation with bottom drag in the
  oceans of Titan and Enceladus}.
\newblock \emph{Icarus}, 281:\penalty0 342--356, 2017.

\bibitem[Hay and Matsuyama(2019)]{Hay2019}
H.~C. Hay and I.~Matsuyama.
\newblock {Nonlinear tidal dissipation in the subsurface oceans of Enceladus
  and other icy satellites}.
\newblock \emph{Icarus}, 319\penalty0 (July 2018):\penalty0 68--85, 2019.

\bibitem[Hemingway and Mittal(2019)]{Hemingway2019}
D.~J. Hemingway and T.~Mittal.
\newblock Enceladus's ice shell structure as a window on internal heat
  production.
\newblock \emph{Icarus}, 2019.

\bibitem[Howett et~al.(2011)Howett, Spencer, Pearl, and Segura]{Howett2011}
C.~J.~A. Howett, J.~R. Spencer, J.~Pearl, and M.~Segura.
\newblock High heat flow from enceladus' south polar region measured using
  10--600 cm−1 cassini/cirs data.
\newblock \emph{Journal of Geophysical Research: Planets}, 116\penalty0 (E3),
  2011.

\bibitem[Iess et~al.(2014)Iess, Stevenson, Parisi, Hemingway, Jacobson, Lunine,
  Nimmo, Armstrong, Asmar, Ducci, and Tortora]{Iess2014}
L.~Iess, D.~J. Stevenson, M.~Parisi, D.~Hemingway, R.~A. Jacobson, J.~I.
  Lunine, F.~Nimmo, J.~W. Armstrong, S.~W. Asmar, M.~Ducci, and P.~Tortora.
\newblock {The Gravity Field and Interior Structure of Enceladus}.
\newblock \emph{Science}, 344\penalty0 (6179):\penalty0 78--80, 2014.

\bibitem[Ivers et~al.(2014)Ivers, Jackson, and Winch]{Ivers2014}
D.~J. Ivers, A.~Jackson, and D.~Winch.
\newblock {Enumeration, Orthogonality and Completeness of the Incompressible
  Coriolis Modes in a Sphere}.
\newblock \emph{Journal of Fluid Mechanics}, pages 468--498, 2014.

\bibitem[Lin and Ogilvie(2018)]{Lin2018}
Y.~Lin and G.~I. Ogilvie.
\newblock {Tidal dissipation in rotating fluid bodies: The presence of a
  magnetic field}.
\newblock \emph{Monthly Notices of the Royal Astronomical Society},
  474\penalty0 (2):\penalty0 1644--1656, 2018.

\bibitem[Matsuyama et~al.(2018)Matsuyama, Beuthe, Hay, Nimmo, and
  Kamata]{Matsuyama2018}
I.~Matsuyama, M.~Beuthe, H.~C. Hay, F.~Nimmo, and S.~Kamata.
\newblock {Ocean tidal heating in icy satellites with solid shells}.
\newblock \emph{Icarus}, 312:\penalty0 208--230, 2018.

\bibitem[Morize et~al.(2010)Morize, {Le Bars}, {Le Gal}, and
  Tilgner]{Morize2010}
C.~Morize, M.~{Le Bars}, P.~{Le Gal}, and A.~Tilgner.
\newblock {Experimental determination of zonal winds driven by tides}.
\newblock \emph{Physical Review Letters}, 104\penalty0 (21):\penalty0 28--31,
  2010.

\bibitem[Nimmo et~al.(2018)Nimmo, Barr, B{\v{e}}hounkov{\'{a}}, and
  Mckinnon]{Nimmo2018}
F.~Nimmo, A.~C. Barr, M.~B{\v{e}}hounkov{\'{a}}, and W.~B. Mckinnon.
\newblock {The thermal and orbital evolution of Enceladus: observational
  constraints and models}.
\newblock \emph{Enceladus and the Icy Moons of Saturn}, pages 79--94, 2018.

\bibitem[Ogilvie(2009)]{Ogilvie2009}
G.~I. Ogilvie.
\newblock {Tidal dissipation in rotating fluid bodies: A simplified model}.
\newblock \emph{Monthly Notices of the Royal Astronomical Society},
  396\penalty0 (2):\penalty0 794--806, 2009.

\bibitem[Ogilvie(2013)]{Ogilvie2013}
G.~I. Ogilvie.
\newblock {Tides in rotating barotropic fluid bodies: The contribution of
  inertial waves and the role of internal structure}.
\newblock \emph{Monthly Notices of the Royal Astronomical Society},
  429\penalty0 (1):\penalty0 613--632, 2013.

\bibitem[Ojakangas and Stevenson(1986)]{Ojakangas1986}
G.~W. Ojakangas and D.~J. Stevenson.
\newblock {Episodic volcanism of tidally heated satellites with application to
  Io}.
\newblock \emph{Icarus}, 66\penalty0 (2):\penalty0 341--358, 1986.

\bibitem[Poincar{\'{e}}(1885)]{Poincare1885}
H.~Poincar{\'{e}}.
\newblock {Sur l'{\'{e}}quilibre d'une masse fluide anim{\'{e}}e d'un mouvement
  de rotation}.
\newblock \emph{Acta Mathematica}, 7\penalty0 (1):\penalty0 259--380, 1885.

\bibitem[Porco et~al.(2006)Porco, Helfenstein, Thomas, Ingersoll, Wisdom, West,
  Neukum, Denk, Wagner, Roatsch, Kieffer, Turtle, McEwen, Johnson, Rathbun,
  Veverka, Wilson, Perry, Spitale, Brahic, Burns, DelGenio, Dones, Murray, and
  Squyres]{Porco2006}
C.~C. Porco, P.~Helfenstein, P.~C. Thomas, A.~P. Ingersoll, J.~Wisdom, R.~West,
  G.~Neukum, T.~Denk, R.~Wagner, T.~Roatsch, S.~Kieffer, E.~Turtle, A.~McEwen,
  T.~V. Johnson, J.~Rathbun, J.~Veverka, D.~Wilson, J.~Perry, J.~Spitale,
  A.~Brahic, J.~A. Burns, A.~D. DelGenio, L.~Dones, C.~D. Murray, and
  S.~Squyres.
\newblock {Cassini observes the active south pole of enceladus}.
\newblock \emph{Science}, 311\penalty0 (5766):\penalty0 1393--1401, 2006.

\bibitem[Rekier et~al.(2018)Rekier, Trinh, Triana, and Dehant]{Rekier2018}
J.~Rekier, A.~Trinh, S.~A. Triana, and V.~Dehant.
\newblock {Inertial modes in Near-spherical geometries}.
\newblock \emph{Geophysical Journal International}, pages 777--793, 2018.

\bibitem[Rieutord and Valdettaro(1997)]{Rieutord1997}
M.~Rieutord and L.~Valdettaro.
\newblock {Inertial waves in a rotating spherical shell}.
\newblock \emph{Journal of Fluid Mechanics}, 341:\penalty0 77--99, 1997.

\bibitem[Rieutord et~al.(2000)Rieutord, Georgeot, and Valdettaro]{Rieutord2000}
M.~Rieutord, B.~Georgeot, and L.~Valdettaro.
\newblock {Inertial waves in a rotating spherical shell: attractors and
  asymptotic spectrum}.
\newblock \emph{Journal of Fluid Mechanics}, 435:\penalty0 42, 2000.

\bibitem[Roberts(2015)]{Roberts2015}
J.~H. Roberts.
\newblock {The fluffy core of Enceladus}.
\newblock \emph{Icarus}, 258:\penalty0 54--66, 2015.

\bibitem[Rovira-Navarro et~al.(2019)Rovira-Navarro, Rieutord, Gerkema, Maas,
  van~der Wal, and Vermeersen]{Rovira-Navarro2019}
M.~Rovira-Navarro, M.~Rieutord, T.~Gerkema, L.~R. Maas, W.~van~der Wal, and
  B.~Vermeersen.
\newblock {Do tidally-generated inertial waves heat the subsurface oceans of
  Europa and Enceladus?}
\newblock \emph{Icarus}, 321:\penalty0 126--140, 2019.

\bibitem[Spencer et~al.(2006)Spencer, Pearl, Segura, Flasar, Mamoutkine,
  Romani, Buratti, Hendrix, Spilker, and Lopes]{Spencer2006}
J.~R. Spencer, J.~C. Pearl, M.~Segura, F.~M. Flasar, A.~Mamoutkine, P.~Romani,
  B.~J. Buratti, A.~R. Hendrix, L.~J. Spilker, and R.~M.~C. Lopes.
\newblock Cassini encounters enceladus: Background and the discovery of a south
  polar hot spot.
\newblock \emph{Science}, 311\penalty0 (5766):\penalty0 1401--1405, 2006.

\bibitem[Thomas et~al.(2016)Thomas, Tajeddine, Tiscareno, Burns, Joseph,
  Loredo, Helfenstein, and Porco]{Thomas2016}
P.~Thomas, R.~Tajeddine, M.~Tiscareno, J.~Burns, J.~Joseph, T.~Loredo,
  P.~Helfenstein, and C.~Porco.
\newblock {Enceladus's measured physical libration requires a global subsurface
  ocean}.
\newblock \emph{Icarus}, 264:\penalty0 37--47, 2016.

\bibitem[Tyler(2014)]{Tyler2014}
R.~Tyler.
\newblock {Comparative estimates of the heat generated by ocean tides on icy
  satellites in the outer solar system}.
\newblock \emph{Icarus}, 243:\penalty0 358--385, 2014.

\bibitem[{Van Hoolst} et~al.(2016){Van Hoolst}, Baland, and
  Trinh]{VanHoolst2016}
T.~{Van Hoolst}, R.~M. Baland, and A.~Trinh.
\newblock {The diurnal libration and interior structure of Enceladus}.
\newblock \emph{Icarus}, 277:\penalty0 311--318, 2016.

\bibitem[Vance et~al.(2018)Vance, Panning, St{\"a}hler, Cammarano, Bills,
  Tobie, Kamata, Kedar, Sotin, Pike, Lorenz, Huang, Jackson, and
  Banerdt]{Vance2018}
S.~D. Vance, M.~P. Panning, S.~St{\"a}hler, F.~Cammarano, B.~G. Bills,
  G.~Tobie, S.~Kamata, S.~Kedar, C.~Sotin, W.~T. Pike, R.~Lorenz, H.-H. Huang,
  J.~M. Jackson, and B.~Banerdt.
\newblock Geophysical investigations of habitability in ice-covered ocean
  worlds.
\newblock \emph{Journal of Geophysical Research: Planets}, 123\penalty0
  (1):\penalty0 180--205, 2018.

\bibitem[Wilson and Kerswell(2018)]{Wilson2018}
A.~Wilson and R.~R. Kerswell.
\newblock {Can libration maintain Enceladus's ocean?}
\newblock \emph{Earth and Planetary Science Letters}, 500:\penalty0 41--46,
  2018.

\end{thebibliography}
